\PassOptionsToPackage{unicode}{hyperref}
\PassOptionsToPackage{hyphens}{url}
\documentclass[
]{article}
\usepackage{lmodern}
\usepackage{amssymb,amsmath}
\usepackage{ifxetex,ifluatex}
\ifnum 0\ifxetex 1\fi\ifluatex 1\fi=0 
\usepackage[T1]{fontenc}
\usepackage[utf8]{inputenc}
\usepackage{textcomp} 
\else 
\usepackage{unicode-math}
\defaultfontfeatures{Scale=MatchLowercase}
\defaultfontfeatures[\rmfamily]{Ligatures=TeX,Scale=1}
\fi
\IfFileExists{upquote.sty}{\usepackage{upquote}}{}
\IfFileExists{microtype.sty}{
	\usepackage[]{microtype}
	\UseMicrotypeSet[protrusion]{basicmath} 
}{}
\makeatletter
\@ifundefined{KOMAClassName}{
	\IfFileExists{parskip.sty}{%
		\usepackage{parskip}
	}{
		\setlength{\parindent}{0pt}
		\setlength{\parskip}{6pt plus 2pt minus 1pt}}
}{
	\KOMAoptions{parskip=half}}
\makeatother
\usepackage{xcolor}
\IfFileExists{xurl.sty}{\usepackage{xurl}}{} 
\IfFileExists{bookmark.sty}{\usepackage{bookmark}}{\usepackage{hyperref}}
\hypersetup{
	hidelinks,
	pdfcreator={LaTeX via pandoc}}
\date{}

\usepackage{graphicx}
\usepackage{hyperref}
\usepackage{pdflscape} 
\usepackage{multirow}
\usepackage{setspace}
\usepackage{lineno}
\usepackage{authblk}
\usepackage{soul}

\usepackage{pdfpages}

\usepackage{fancyhdr}

\usepackage{booktabs}

\usepackage{longtable}

\title{A Graph Theory approach to assess nature’s contribution to people at a global scale\footnote{© 2020. This manuscript version is made available under the CC-BY-NC-ND 4.0 license \href{http://creativecommons.org/licenses/by-nc-nd/4.0/}{http://creativecommons.org/licenses/by-nc-nd/4.0/}}}

\author[1]{Silvia de Juan Mohan}

\author[2]{Andres Ospina-Alvarez\footnote{Corresponding author: Andrés Ospina-Alvarez, email: aospina.co@me.com; address: Spanish Scientific Research Council, Mediterranean Institute for Advanced Studies (IMEDEA-CSIC/UIB), C/ Miquel Marques 21, CP 07190 Esporles, Balearic Islands, Spain.}}

\author[3, 4]{Sebastián Villasante}

\author[2]{Ana Ruiz-Frau}

\affil[1]{Marine Science Institute (ICM-CSIC), Passeig Marítim de la
	Barceloneta, 37-49, CP 08003 Barcelona, Catalunya, Spain.}
\affil[2]{Mediterranean Institute for Advanced Studies (IMEDEA-CSIC/UIB), C/
	Miquel Marques 21, CP 07190 Esporles, Balearic Islands, Spain.}
\affil[3]{Faculty of Economics and Business Administration,
	University of Santiago de Compostela, Av~Burgo das Nacions~s/n, 15782
	Santiago de Compostela, A Coruña, Spain.}
\affil[4]{Campus Do Mar, International Campus of Excellence,
	Spain.}

\begin{document}
	\maketitle
	
	\textbf{Running page head:} Wide-scale assessment of CES using Graph Theory
	
	\begin{abstract}
		
		Cultural Ecosystem Services (CES) assessment at
		large scales is crucial in marine ecosystems as they reflect key
		physical and cognitive interactions between humans and nature. The
		analysis of social media data with graph theory is a promising approach
		to provide global information on users' perceptions for different marine
		ecosystems. Fourteen areas were selected to illustrate the use of graph
		theory on social media data. The selected areas, known to protect key
		recreational, educational and heritage attributes of marine ecosystems,
		were investigated to identify variability in users' preferences.
		Instagram data (i.e., hashtags associated to photos) was extracted for
		each area allowing an in-depth assessment of the CES most appreciated by
		the users. Hashtags were analysed using network centrality measures to
		identify clusters of words, aspects not normally captured by traditional
		photo content analysis. The emergent properties of networks of hashtags
		were explored to characterise visitors' preferences (e.g., cultural
		heritage or nature appreciation), activities (e.g., diving or hiking),
		preferred habitats and species (e.g. forest, beach, penguins), and
		feelings (e.g., happiness or place {identity). }Network analysis on
		Instagram hashtags allowed delineating the users' discourse around a
		natural area, which provides crucial information for effective
		management of popular natural spaces for people.{~}
		
	\end{abstract}


\section{Introduction}

Marine and coastal areas are extremely important for peoples' wellbeing
and yet, management plans rarely consider Cultural Ecosystem Services
(CES) in their formulation {(Chen et al., 2020; Rodrigues et al.,
2017)}. CES are recognized as a main pillar in ecosystem services
frameworks {(Liquete et al., 2013)}, however, CES are the most
challenging group of ecosystem services to study, principally due to
their intangible and subjective nature {(Daniel et al., 2012; Kirchhoff,
2012)}. Additionally, research targeted at marine CES has mostly focused
on the economic valuation of recreational activities, tourism, or
seascape scenic beauty {(Milcu et al., 2013; Teoh et al., 2019)},
setting aside the non-material benefits people obtain from ecosystems
that have symbolic, cultural or intellectual significance {(Haines-Young
and Potschin, 2010)}. During the past decade, the understanding of CES
has evolved to acknowledge the importance of the relationship between
people and nature, as CES are the outcome of the interaction between
these two ecosystem components {(Chan et al., 2012)}. Despite the
challenges associated to CES assessment, current management schemes
should incorporate the multi-dimensional CES valuation at scales
relevant for management, particularly in marine and coastal areas where
there is high economic and cultural dependency on marine ecosystems
{(Russell et al., 2013)}.{~}

The monitoring of CES at large spatial scales is particularly difficult
because, among others, they often have been based on methods developed
for small spatial scales. Field survey methods have been generally used
{(Gosal et al., 2019)}, including interviews, face-to-face
questionnaires and participatory mapping (e.g., {Klain et al., 2014;
Oteros-Rozas et al., 2014; Plieninger et al., 2013; Ruiz-Frau et al.,
2011)}. These studies generally focus on local scales {(Clemente et al.,
2019)}, whereas management generally needs information at regional
scales. In this context, new methodological approaches are needed to
assess the multiple cultural values provided by marine ecosystems at
scales larger than the local case study. There are several studies that
adopt global approaches {(Chen et al., 2020; Costanza et al., 2014)},
but generally these imply low cost-effectiveness that is a requirement
for widely adopted assessments.{~}

The volume of information uploaded to online social media platforms,
like Instagram or Flickr, can provide an important source of information
to assess peoples' preferences and values through a cost-effective
approach {(Clemente et al., 2019; Retka et al., 2019; Vaz et al.,
2019)}. Social media platforms continuously store information people
upload from any location in the planet. These sites are used for
socializing and communicating, frequently focusing on recreational
activities, including tourism {(Figueroa-Alfaro and Tang, 2016)}. As
part of the information uploaded, people often express their perceptions
and feelings about places {(Hale et al., 2019)}, including natural
spaces. In the internet era, there are many social network platforms
with millions of users that are an important source of big data (Liu et
al., 2014). In the quest to avoid the time-consuming nature of field
surveys and to identify alternative methods, there has been an
increasing number of scientific studies that use social media to assess
CES {(Figueroa-Alfaro and Tang, 2016; Gosal et al., 2019; Oteros-Rozas
et al., 2018)}. These studies have generally proved to be comparable to
traditional surveys (e.g., {Hausmann et al., 2018)}.

{Social media data allows an indirect assessment of peoples' perceptions
and preferences. It provides large sample sizes and data is available at
global scales. }The predominant approach in social media data analysis
generally relies on photo content assessment (but see {Geboers and Van
De Wiele, 2020)}. The context and content of the photographs is
classified into CES categories based on the presence or absence of
specific elements in the photos, such as views of flora and fauna,
historical buildings, or touristic infrastructure and facilities
{(Ghermandi et al., 2020)}. Most works conducted up to date tend to use
Flickr as source of data and analyse the photo content manually {(Jeawak
et al., 2017)} or through automatic identification {(Lee et al., 2019)}.
An advantage associated to the use of Flickr is the availability of
geolocalised photos. The downside of this platform, however, is a
relatively low number of users, its decreasing popularity, and a photo
content strongly dominated by biodiversity and landscape aspects
{(Oteros-Rozas et al., 2018)}, limiting the scope of the CES assessment.
Conversely, Instagram is generally used to post photographs and thoughts
in real-time often related to activities or social recreation, but also
to culture and wildlife appreciation {(Ruiz-Frau et al., 2020)}.
Instagram users' demographics, however, are dominated by younger
generations {(Hausmann et al., 2018)}. An advantage associated to
Instagram is the frequent inclusion of hashtags as part of the photo
post. These hashtags are used as keywords to mark messages or form
conversations, and thus they provide an additional way to connect visual
content (i.e., photos) and semantically related words to a discourse.
The user-generated hashtags provide a great opportunity to analyse the
discourse linked to the posted photos and minimize the subjectivity and
low-cost effectiveness associated to photo content analysis.{~}

Recent developments in the analysis of social media data have applied
graph theory to the analysis of hashtags associated to posts providing a
promising approach for the remote assessment of CES relying on social
media data {(Ruiz-Frau et al., 2020)}. Initial results indicate that the
use of this cost-effective approach reveals, besides the more tangible
set of CES such as recreational activities, a set of intangible CES
aspects such as relational values {(Ruiz-Frau et al., 2020)}, providing
a more encompassing view of CES provision. The application of this
approach, however, has been so far limited to the regional scale while
it offers an untapped potential to be applied at a global scale,
providing comparative information on the type of CES contribution that
marine and coastal areas make around the world.{~}

Graph theory, as the mathematical study of the interaction of a system
of connected elements, is a suitable approach for analysing user
behaviour in social networks. It provides a simplified and quantitative
view of the multiple factors involved in the exchange among system
elements (Freeman, 1979). A system of connected elements can be defined
as a network, also called a graph. In a network of keywords posted with
the photos, graph theory provides insights into the system properties
and identifies critical nodes with high centrality (i.e., words
connected to many other words) or clusters of well-connected nodes
(Maiya and Berger-Wolf, 2010; Roth and Cointet, 2010; Topirceanu et al.,
2018). In this study, the working hypothesis was that data extracted
from online social networks and analysed by calculating different
measures of centrality from graph theory can be used to understand
peoples' preferences for nature and nature-based experiences in marine
and coastal areas worldwide. The hypothesis was tested in 14 marine and
coastal areas that are expected to provide a wide diversity of CES and
the methodology was applied to determine whether different areas of the
world were delivering different arrays of CES. This approach is expected
to contribute to cost-efficient large-scale assessments of the
contribution of marine and coastal areas to society well-being.{~}

\section{Materials and Methods}

\subsection{Case studies}

In order to encompass a wide diversity of marine and coastal ecosystems
across regions, and the potential diversity of CES provided by these
areas, 14 marine and coastal areas were chosen for the study (Table A1).
These areas span over the 12 marine realms established by (Spalding et
al., 2007) and are expected to provide a wide diversity of CES (e.g.,
recreation, cultural heritage, nature and wildlife observation) and to
be visited by a wide diversity of users. The areas chosen had to comply
with two criteria: 1) the area had to be sufficiently popular to contain
enough data for the analysis; 2) the name of the area had to be
sufficiently characteristic to provide a unique identifier within
Instagram (Table 1). The adoption of these criteria meant that no area
in the Artic (marine realm 1) could be identified with enough social
media data. Some of the realms established by Spalding et al. (2007) are
too broad to capture existent variability across systems (e.g.,
temperate Northern Atlantic and Mediterranean Sea), when the authors
considered this was the case, more than one study areas were selected to
capture this variability (Fig. 1).

\begin{longtable}[]{@{}llll@{}}
\caption{Case study name, location, query and number of downloaded for the study}
\label{tab:table1}\\
\hline
\toprule
\endhead
\begin{minipage}[t]{0.22\columnwidth}\raggedright
\textbf{Case study}\strut
\end{minipage} & \begin{minipage}[t]{0.22\columnwidth}\raggedright
\textbf{Location}\strut
\end{minipage} & \begin{minipage}[t]{0.32\columnwidth}\raggedright
\textbf{Query}\strut
\end{minipage} & \begin{minipage}[t]{0.22\columnwidth}\raggedright
\textbf{Number of posts}\strut
\end{minipage}\tabularnewline
\begin{minipage}[t]{0.22\columnwidth}\raggedright
Galapagos\strut
\end{minipage} & \begin{minipage}[t]{0.22\columnwidth}\raggedright
Ecuador\strut
\end{minipage} & \begin{minipage}[t]{0.32\columnwidth}\raggedright
\#galapagos\strut
\end{minipage} & \begin{minipage}[t]{0.22\columnwidth}\raggedright
10,000\strut
\end{minipage}\tabularnewline
\begin{minipage}[t]{0.22\columnwidth}\raggedright
Glacier Bay\strut
\end{minipage} & \begin{minipage}[t]{0.22\columnwidth}\raggedright
Alaska\strut
\end{minipage} & \begin{minipage}[t]{0.32\columnwidth}\raggedright
\#glacierbayalaska\strut
\end{minipage} & \begin{minipage}[t]{0.22\columnwidth}\raggedright
1,811\strut
\end{minipage}\tabularnewline
\begin{minipage}[t]{0.22\columnwidth}\raggedright
Great Barrier Reef\strut
\end{minipage} & \begin{minipage}[t]{0.22\columnwidth}\raggedright
Australia\strut
\end{minipage} & \begin{minipage}[t]{0.32\columnwidth}\raggedright
\#greatbarrierreef\strut
\end{minipage} & \begin{minipage}[t]{0.22\columnwidth}\raggedright
9,960\strut
\end{minipage}\tabularnewline
\begin{minipage}[t]{0.22\columnwidth}\raggedright
Isole Egadi\strut
\end{minipage} & \begin{minipage}[t]{0.22\columnwidth}\raggedright
Italy\strut
\end{minipage} & \begin{minipage}[t]{0.32\columnwidth}\raggedright
\#isoleegadi\strut
\end{minipage} & \begin{minipage}[t]{0.22\columnwidth}\raggedright
9,969\strut
\end{minipage}\tabularnewline
\begin{minipage}[t]{0.22\columnwidth}\raggedright
Macquarie Island\strut
\end{minipage} & \begin{minipage}[t]{0.22\columnwidth}\raggedright
Australia\strut
\end{minipage} & \begin{minipage}[t]{0.32\columnwidth}\raggedright
\#macquarieisland\strut
\end{minipage} & \begin{minipage}[t]{0.22\columnwidth}\raggedright
1,430\strut
\end{minipage}\tabularnewline
\begin{minipage}[t]{0.22\columnwidth}\raggedright
Peninsula Valdez\strut
\end{minipage} & \begin{minipage}[t]{0.22\columnwidth}\raggedright
Argentina\strut
\end{minipage} & \begin{minipage}[t]{0.32\columnwidth}\raggedright
\#peninsulavaldes\strut
\end{minipage} & \begin{minipage}[t]{0.22\columnwidth}\raggedright
9,971\strut
\end{minipage}\tabularnewline
\begin{minipage}[t]{0.22\columnwidth}\raggedright
Easter Island\strut
\end{minipage} & \begin{minipage}[t]{0.22\columnwidth}\raggedright
Chile\strut
\end{minipage} & \begin{minipage}[t]{0.32\columnwidth}\raggedright
\#easterisland, \#rapanui, \#isladepascua\strut
\end{minipage} & \begin{minipage}[t]{0.22\columnwidth}\raggedright
10,000\strut
\end{minipage}\tabularnewline
\begin{minipage}[t]{0.22\columnwidth}\raggedright
Sandwich Harbour\strut
\end{minipage} & \begin{minipage}[t]{0.22\columnwidth}\raggedright
Namibia\strut
\end{minipage} & \begin{minipage}[t]{0.32\columnwidth}\raggedright
\#sandwichharbour\strut
\end{minipage} & \begin{minipage}[t]{0.22\columnwidth}\raggedright
2,807\strut
\end{minipage}\tabularnewline
\begin{minipage}[t]{0.22\columnwidth}\raggedright
Skomer\strut
\end{minipage} & \begin{minipage}[t]{0.22\columnwidth}\raggedright
United Kingdom\strut
\end{minipage} & \begin{minipage}[t]{0.32\columnwidth}\raggedright
\#skomer\strut
\end{minipage} & \begin{minipage}[t]{0.22\columnwidth}\raggedright
4,911\strut
\end{minipage}\tabularnewline
\begin{minipage}[t]{0.22\columnwidth}\raggedright
Tawharanui\strut
\end{minipage} & \begin{minipage}[t]{0.22\columnwidth}\raggedright
New Zealand\strut
\end{minipage} & \begin{minipage}[t]{0.32\columnwidth}\raggedright
\#tawharanui\strut
\end{minipage} & \begin{minipage}[t]{0.22\columnwidth}\raggedright
6,832\strut
\end{minipage}\tabularnewline
\begin{minipage}[t]{0.22\columnwidth}\raggedright
Tayrona\strut
\end{minipage} & \begin{minipage}[t]{0.22\columnwidth}\raggedright
Colombia\strut
\end{minipage} & \begin{minipage}[t]{0.32\columnwidth}\raggedright
\#tayrona\strut
\end{minipage} & \begin{minipage}[t]{0.22\columnwidth}\raggedright
10,000\strut
\end{minipage}\tabularnewline
\begin{minipage}[t]{0.22\columnwidth}\raggedright
Togean Island\strut
\end{minipage} & \begin{minipage}[t]{0.22\columnwidth}\raggedright
Indonesia\strut
\end{minipage} & \begin{minipage}[t]{0.32\columnwidth}\raggedright
\#togeanisland\strut
\end{minipage} & \begin{minipage}[t]{0.22\columnwidth}\raggedright
9,467\strut
\end{minipage}\tabularnewline
\begin{minipage}[t]{0.22\columnwidth}\raggedright
Vamizi\strut
\end{minipage} & \begin{minipage}[t]{0.22\columnwidth}\raggedright
Mozambique\strut
\end{minipage} & \begin{minipage}[t]{0.32\columnwidth}\raggedright
\#vamizi\strut
\end{minipage} & \begin{minipage}[t]{0.22\columnwidth}\raggedright
1,367\strut
\end{minipage}\tabularnewline
\begin{minipage}[t]{0.22\columnwidth}\raggedright
Ytrehvaler\strut
\end{minipage} & \begin{minipage}[t]{0.22\columnwidth}\raggedright
Norway\strut
\end{minipage} & \begin{minipage}[t]{0.32\columnwidth}\raggedright
\#ytrehvalernasjonalpark\strut
\end{minipage} & \begin{minipage}[t]{0.22\columnwidth}\raggedright
1,019\strut
\end{minipage}\tabularnewline
\bottomrule
\end{longtable}

\begin{figure}
	\centering
	\includegraphics[width=1\linewidth]{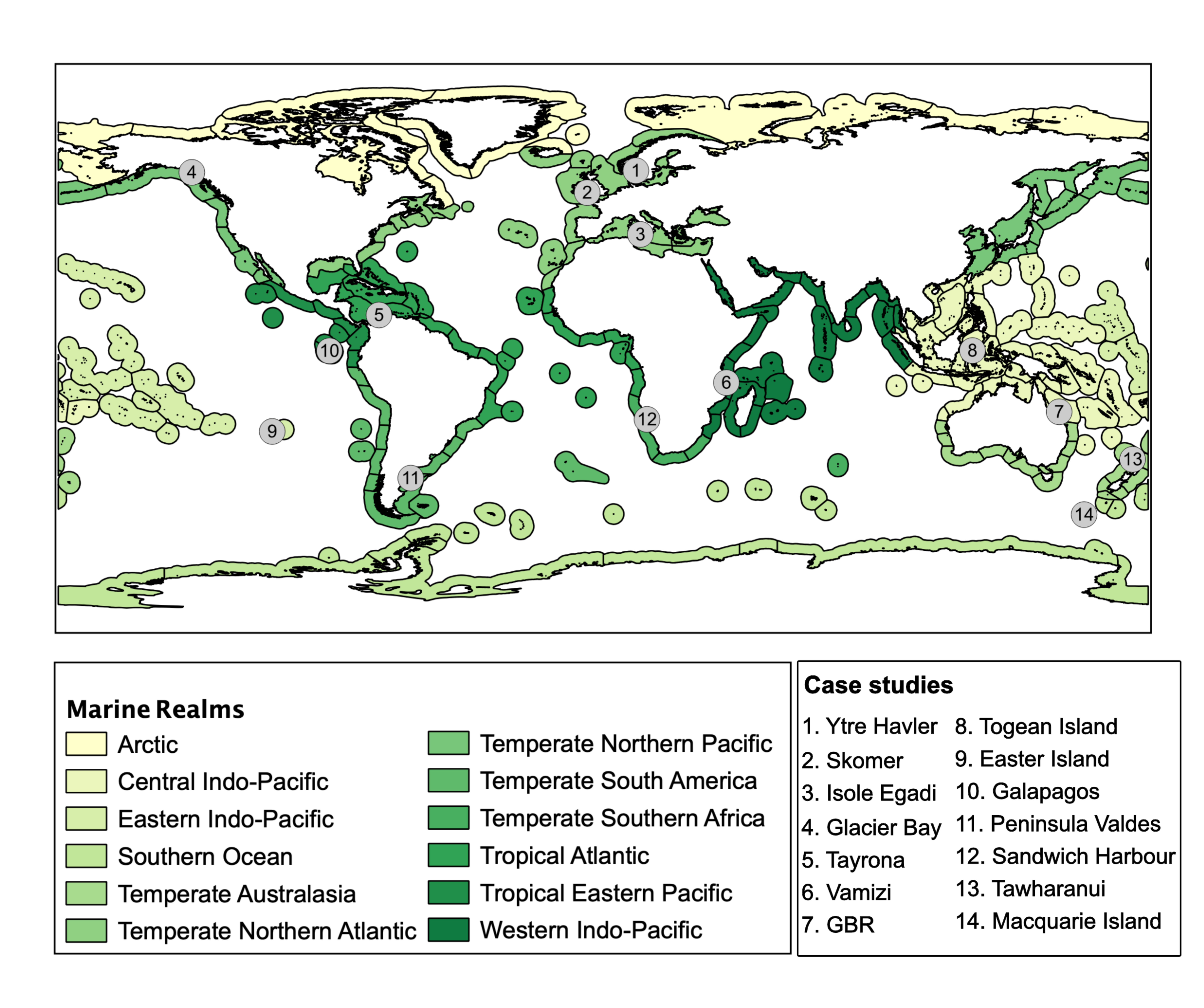}
	\caption{The fourteen case studies selected across the twelve marine realms proposed by Spalding et al (2007)}
	\label{fig:fig1}
\end{figure}

\subsection{Social media data}

Data collection and analysis were carried out according to the
methodology established in Ruiz-Frau et al. (2020). Instagram posts were
downloaded through the Application Programming Interface (API), with a
specific development for the R language and environment for statistical
computing version 3.6.0, released 2019-04-26 (R Development Core Team,
2009). The Instagram API is suitable for a hashtag-based data extraction
and, for each case study, a search query was executed (Table 1). Query
terms were based on the hashtags of the geographical name of the study
areas; therefore, the post download was related to a specific query, or
name of the study area (e.g., Galapagos), with all downloaded posts
including this query. We downloaded 10,000 posts per case study in June
2019. No specific period was defined. The data download started with the
most recent post and was followed by the previous post until reaching
the cut-off (i.e., 10,000). Some of the case studies had fewer than the
established threshold (i.e., 10,000 posts), in such cases we downloaded
all available posts (Table 1). Query search was limited to English, the
most common language amongst tourists; this, however, might have
overlooked posts where the name of the place was in a different
language. For most marine areas, this was considered irrelevant as the
name of the place is not translated to other languages (e.g., Tayrona,
Vamizi, Skomer). In some of the cases the name of the place could appear
in a variety of languages (e.g., Great Barrier Reef), however the use of
non-English place hashtags as queries generally retrieved a
significantly lower number of posts (e.g. Gran Barrera de Coral in
Spanish with 1,900 posts, or Grand Barrière de Corail in French with 14
posts, while Great Barrier Reef had over 10,000 posts). In the specific
case of Easter Island, we observed that the use of three particular
queries was linked to a high number of posts: Easter Island and the
local name Rapanui had over 10,000 posts each, and Isla de Pascua in
Spanish had 8,700 posts. In this case, three separate posts' downloads
were performed, and data were merged for subsequent analysis.{~}

Downloaded posts for each case study were stored locally and datasets
were filtered and cleaned in order to retain only relevant information
for further analysis (Di Minin et al., 2018; Varol et al., 2017). Posts
often contain non-relevant information as social media platforms are
frequently used as marketing and advertisement tools to reach a wider
public and often bots (automated data generating algorithms and
advertisements) are used to created large volumes of automated posts. In
our case, irrelevant posts, mostly related to advertisement (e.g., posts
related with a trading mark named Galapagos or Rapanui), were discarded
from the analysis. Discarded photos were done through excluding posts
with a specific hashtag (e.g., \#chocolate, frequently linked with
\#rapanui due to a trademark) or a specific user (i.e., those users
identified as posting marketing). Dataset cleaning also consisted in
merging similar words (e.g., \#travelgram, \#instatravel, \#igtravel)
and misspellings (e.g., \#travel, \#travell). Highly frequent
non-English words were translated to English (e.g., \#statue,
\#steinfigure, \#estatua; for Easter Island statue in different
languages) to homogenise the network language and avoid numerous
duplicates. However, in some networks with a prevalence of non-English
language (e.g., Ytrehvaler in Norwegian) words were not translated to
English to capture users' characteristics.{~}

\subsection{Graph theory}
The analysis of networks using graph theory can be described as the
analysis of existing relationships between the different elements
contained in a network. The term \emph{vertex} is used to describe the
elements in a network, while the term \emph{edge} is used to refer to
the connections between the different vertices in a network. In our
case, vertices are represented by hashtags, while edges illustrate the
connections between hashtags (e.g., the hashtags included in the same
posts and the frequency of those connections). To assess relationships
between hashtags and identify emerging themes within the networks, we
used and expanded the centrality measures and community algorithms
established in Ruiz-Frau et al. (2020).{~}

The concept of \emph{centrality} is a commonly used metric in the
analysis of networks. The identification of important, or
\emph{central,} vertices in a network is a key aspect in the definition
and description of networks (Bodin et al., 2006). However, there are
multiple interpretations of what makes a vertex important and there are
therefore many measures of centrality (Freeman, 1979). In this work, we
will focus on the use of \emph{betweenness} and \emph{eigenvector
centrality} to illustrate and interpret the structure of the social
networks.{~}

Eigenvector centrality {measure }(Bonacich, 2007){ takes into
consideration not only how many connections a vertex has, but also the
centrality of the vertices that it is connected to.} Eigenvector
centrality, hereafter eigenvector, ranks higher those vertices that are
connected to important neighbours, i.e., other vertices that are
connected to many other vertices. It is a measure of the influence of a
vertex in a network. In our study context, hashtags with high
eigenvector values are high frequency hashtags that in turn are
connected to other high frequency hashtags, and so on. Therefore, this
measure allows identifying those hashtags that are frequently posted
with other hashtags also frequently posted, and it can be interpreted as
the pairs or groups of features more frequently related to the case
study by the users.{~}

Betweenness {centrality }(Freeman, 1979){ is a measure of the influence
of a vertex over the flow of information between every pair of vertices
under the assumption that information primarily flows over the shortest
paths between them. }Betweenness centrality indicates nodes that have a
high probability of having routes that connect them to other nodes in
the network. Alternatively, it indicates nodes in an intermediate
position between groups of very well-connected neighbouring nodes. Nodes
with high betweenness centralities have been termed \emph{bottlenecks}
or \emph{bridges} and they prevent the fragmentation of the network.
Similarly, edge betweenness centrality is defined as the number of the
shortest paths that go through an edge in a graph or network (Girvan and
Newman, 2002). Each edge in the network can be associated with an edge
betweenness centrality value. An edge with a high edge betweenness
centrality score represents a bridge-like connector between two parts of
a network, the removal of which may affect the communication between
many pairs of nodes through the shortest paths between them. In our
context, {betweenness centrality (hereafter betweenness) and edge
betweenness centrality (hereafter edge betweenness) provide information
on hashtags and links between hashtags that are essential to structure
the network in sub-communities; the removal of those links would
fragment the network and disconnect the hashtags that have higher
betweenness.}{ }{High betweenness hashtags are those that appear in a
large number of posts and represent concepts that people often identify
as ideal descriptors of a network. Edge betweenness evidences those
edges that connect the most frequent hashtags with other less frequent
hashtags. Therefore, they might show the parallel or additional
discourse to the main discourse of the users, allowing to identify less
frequent activities or perceptions but equally important to understand
the network as a whole.}

\subsection{Data analysis}
To illustrate the most relevant information contained as part of the
10,000 posts for each of the 14 areas, we selected the 150 most frequent
hashtags from each dataset in order to create the network graph. The
first 150 hashtags (frequency \textgreater{} 1.5\% when 10,000 posts are
retrieved) had a probability of more than 90\% of occurring with any
other of the first 150 hashtags in the same post. Therefore, this
criterion was used to create networks with great cohesion and
connectedness, representing a dominant discourse in relation to the area
in question. Network graphs were delineated using eigenvector,
betweenness and edge betweenness as centrality measures. These metrics
were selected as they provided a visualisation of the central hashtags
in the social media discourse, as well as peripherical hashtags that
where nevertheless often related with these central hashtags; therefore,
these provide information on the periphery of the dominant discourse.{~}

In order to find emerging patterns within the 14 case study networks,
hashtags were assigned to communities through the use of Fast-Greedy
community algorithm (Ruiz-Frau et al., 2020). Fast-Greedy algorithm
makes the best choice at each small step in the hope that each of these
small steps will lead to a globally optimal solution (Newman, 2004).
Relevant hashtag communities based on Fast-Greedy algorithm were
assessed to provide a detailed assessment of the type of CES provided by
each of the case studies.

In order to visualise potential similarities in the social media
discourse across the 14 case studies, all the data was merged, and the
1400 most frequent hashtags pairs were retained for analysis in a single
network graph. Similar to what was previously described for the
individual networks, these 1400 pairs of hashtags accounted over 90\% of
the linkages between hashtags, representing the dominant discourse on
CES across the 14 areas. Eigenvector centrality was used as the measure
of vertex influence in the network and connections were represented with
a backbone layout (Brandes and Wagner, 2004). This layout has proven
effective to illustrate networks with most vertices in a central
position that result in high overlap in large networks (Nocaj et al.,
2015).

To conduct the analysis, we used the open source graphics manipulation
software \emph{igraph{~}} (Csardi and Nepusz, 2006) to obtain the centrality measures and
communities aggregations. Graphics and figures were generated using the
visualization software \emph{ggraph} and \emph{ggtree}. All of the above
software can be used as extension packages of the R language and
environment for statistical computing (R Development Core Team, 2009)
freely available online.{~}

\section{Results}
\subsection{Network centrality measures}

Results indicated that network graphs captured information on distinct
types of CES, for example those based on wildlife and nature; heritage;
or beach tourism. In areas such as Galapagos, popular hashtags were
\emph{nature}, \emph{wildlife}, \emph{photography}, \emph{travel} and
\emph{adventure}, evidencing a preference for wildlife and nature based
CES. In this area, betweenness evidenced the connections between the
most frequent hashtags group with other hashtags like \emph{waves},
\emph{crab}, \emph{endemic}, \emph{evolution} and \emph{happy}, and
provided information on the discourse of Galapagos' visitors (Fig. 2).
Other areas providing wildlife and nature based CES were Skomer nature
reserve, characterised by the hashtags \emph{birds} (particularly
\emph{Puffin}), \emph{nature} and \emph{wildlife photography}, and
Península Valdés, characterized by many locality names and by fauna,
with the hashtags' \emph{wildlife}, \emph{whales} and \emph{nature}
holding high eigenvector and funnelling most connections to other
hashtags and providing a full picture of the post (e.g., \emph{wind},
\emph{hiking}, \emph{relax}). Three networks, Sandwich Harbour, Glacier
Bay and Macquarie Island also included popular hashtags related with
\emph{nature}, \emph{wildlife} and \emph{photography}; however, most
hashtags had low betweenness and edge betweenness limiting the diversity
of the posts (all network graphs included in Appendix, Fig. A1).{~}

Regarding cultural heritage, Easter Island was characterised by popular
hashtags related with Easter Island stone statues and with travel, and
edge betweenness evidenced a diversity of peripherical nodes that
describe other cultural elements, like \emph{design}, \emph{music} and
\emph{food.} Other areas reflected cultural identity by the frequent
post of local names (e.g., Ytrehvaler), of words related with the
country's identity (e.g., Isole Egadi) and positive feelings about this
identity (e.g., Tawharanui). In Tayrona National Park network, the full
discourse identified cultural identity like \emph{Kogui} (indigenous
village) linked with the popular posts. However, the most frequent
hashtags in Tayrona network, and also in Tawharanui and Isole Egadi,
were related with \emph{beach}, \emph{nature} and \emph{summer}. In some
of these networks, like Isole Egadi and Ytrehvaler, locality names are
frequently posted, allowing to identify connections between places and
activities, wildlife or natural structures.{~}

\begin{figure}
	\centering
	\includegraphics[width=1\linewidth]{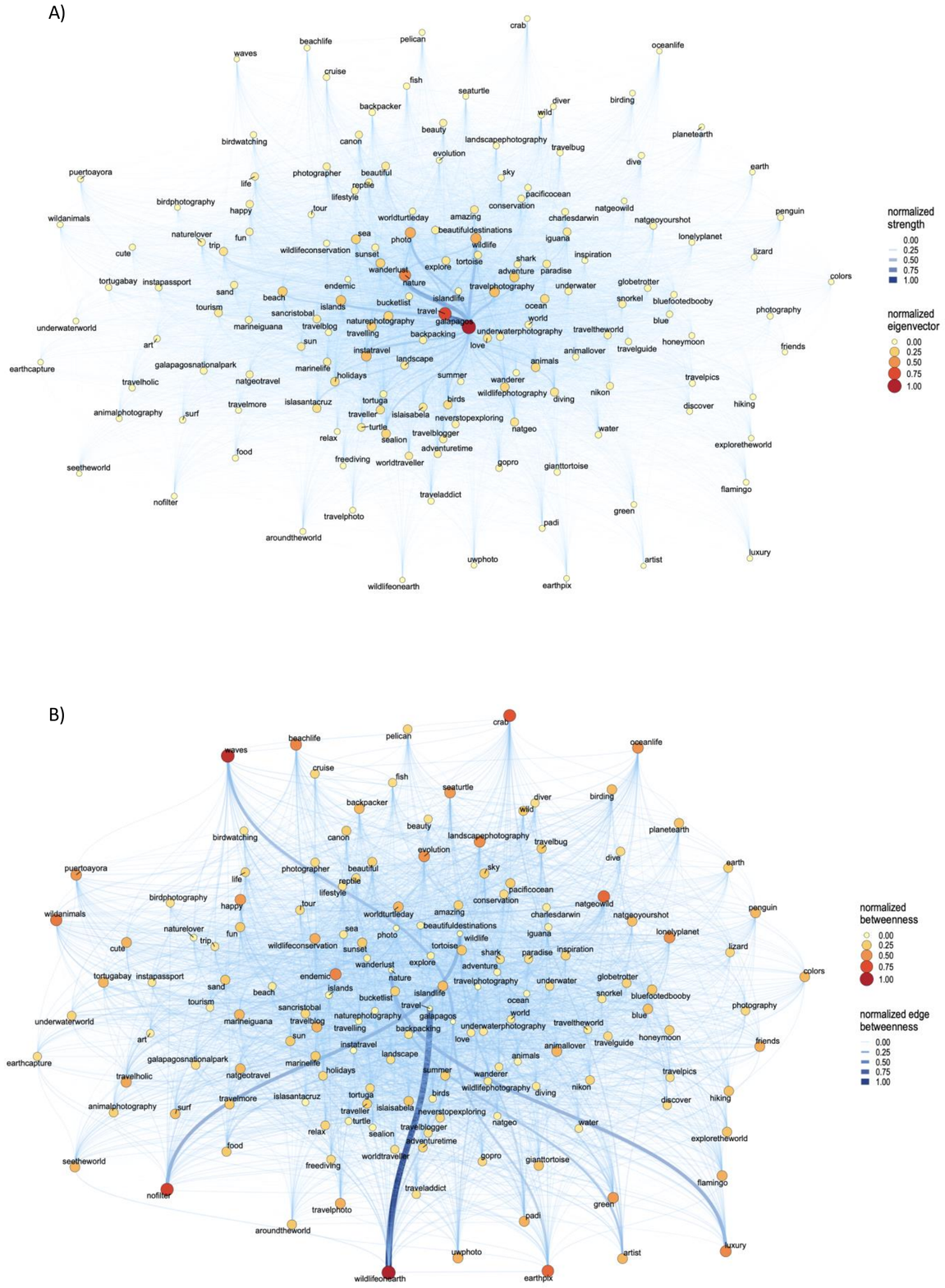}
	\caption{Example of network graphs in Galapagos case study. In plot A) vertex size represents the Eigenvector centrality and edges represent strength. In plot B) vertex size represents Betweenness centrality and edges represent Edge betweenness.}
	\label{fig:fig2}
\end{figure}

A group of areas were appreciated by their underwater ecosystems. For
Great Barrier Reef, popular hashtags were related with the coral reef:
\emph{ocean}, \emph{diving}, \emph{underwater photography},
\emph{travel}, \emph{nature}, \emph{coral} and \emph{reef}; whereas
betweenness highlighted a set of hashtags related with conservation:
\emph{science}, \emph{sustainability}, \emph{save the reef}, \emph{4
ocean} (Fig. 3). In Toguean Island network, the frequent hashtags
\emph{beach}, \emph{wonderful} and \emph{charming} are connected to
peripherical hashtags related with the sea (e.g., \emph{sea life},
\emph{diving}). In Vamizi, popular hashtags were related with tourism,
\emph{private island}, \emph{travel}, \emph{luxury travel}, and were
connected to less frequent hashtags linked to the sea, including
recreational fisheries.{~}

\begin{figure}
	\centering
	\includegraphics[width=1\linewidth]{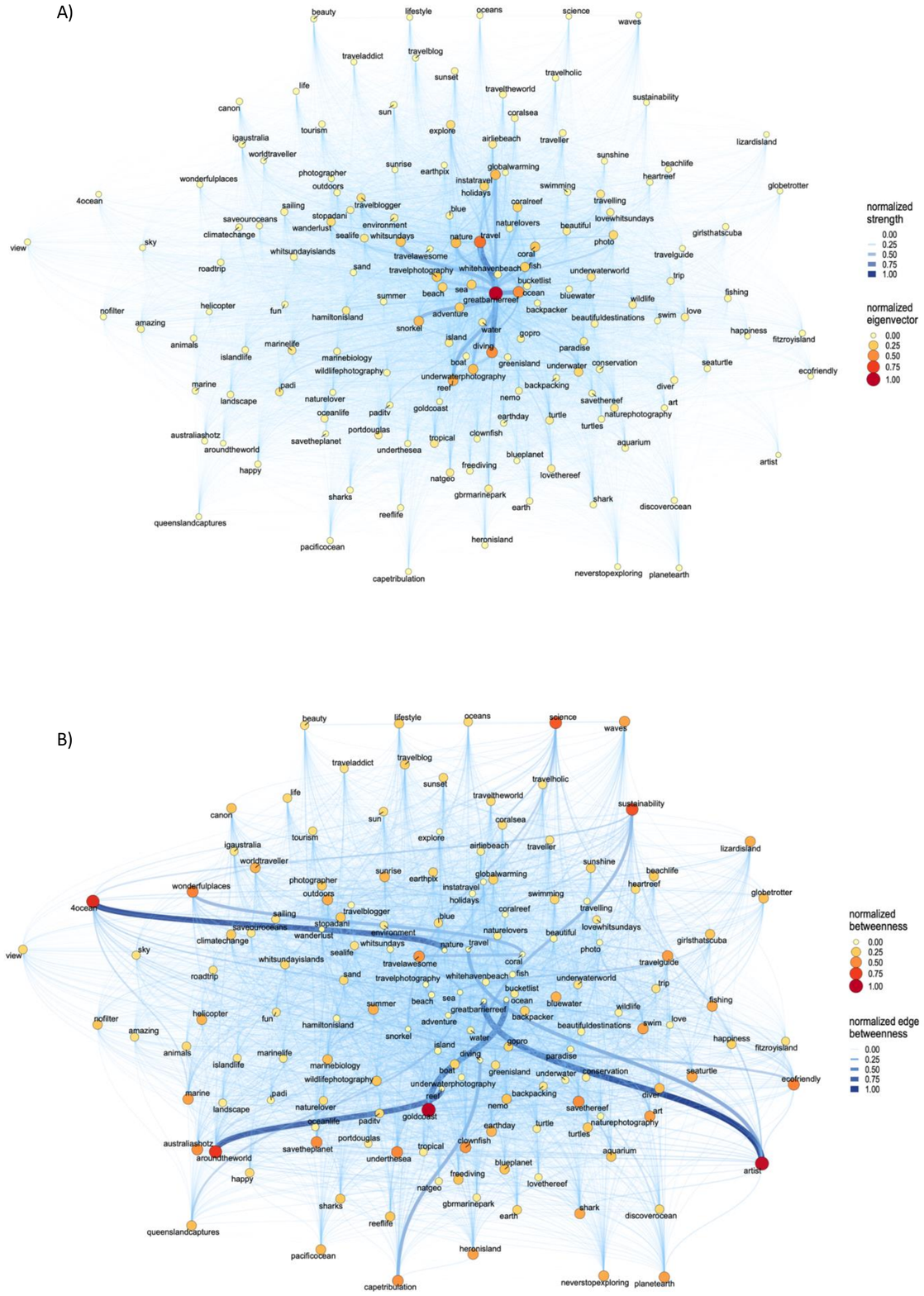}
	\caption{Example of network graphs in Great Barrier Reef case study. In plot A) vertex size represents the Eigenvector centrality and edges represent strength. In plot B) vertex size represents Betweenness centrality and edges represent Edge betweenness.}
	\label{fig:fig3}
\end{figure}

\subsection{Network communities}

The division of hashtags in communities allows for a more detailed
exploration of the words included in the 150 most frequent hashtags
selection, independently of their centrality measures, and allowed a
categorisation of hashtags within CES classes in each area (Table 2).
Hashtags were grouped in 3 to 5 communities, with some communities
relatively constant across case studies, e.g., aesthetics, wildlife and
nature appreciation (Fig. 4, Fig. A2).{~}

In some of the areas, the communities were diverse in hashtag
composition, for example, in Galapagos, \emph{wildlife} (and related
words) was distinctive of several communities, but other communities
were characterised by different concepts: \emph{beach}, \emph{holidays},
\emph{happiness}, \emph{snorkelling} and \emph{diving}. In Easter Island
network the hashtags related with the stone statues and cultural
heritage characterise one community, while the other communities include
a diversity of hashtags classified under adventure, nature, underwater
recreational activities. Tayrona (Fig. 4) is also a diverse network with
one community characterised by hashtags like \emph{beach},
\emph{summer}, \emph{happiness} (wellbeing), but other communities
containing a diversity of hashtags like \emph{forest}, \emph{hiking},
\emph{indigenous} and \emph{wildlife} (classified in recreational,
cultural heritage, nature and aesthetics; Table 2).

In some areas, the communities were not so diverse, but provided
additional information on the posts. For example, in MacQuarie Island
the communities highlighted iconic fauna, including several penguin
species, and biodiversity conservation. In several areas, network
communities informed of the iconic fauna and specific places: puffins
and other bird species in Skomer; southern right whale, sealions and
penguins in Península Valdés; glaciers and mountains in Glacier bay
(Fig. 4); desert and dunes in Sandwich harbour. Finally, Ytrehvaler is a
network characterised by many local names (in Norwegian), evidencing a
national tourism, and hashtags related with scenery.{~}

\begin{figure}
	\centering
	\includegraphics[width=1\linewidth]{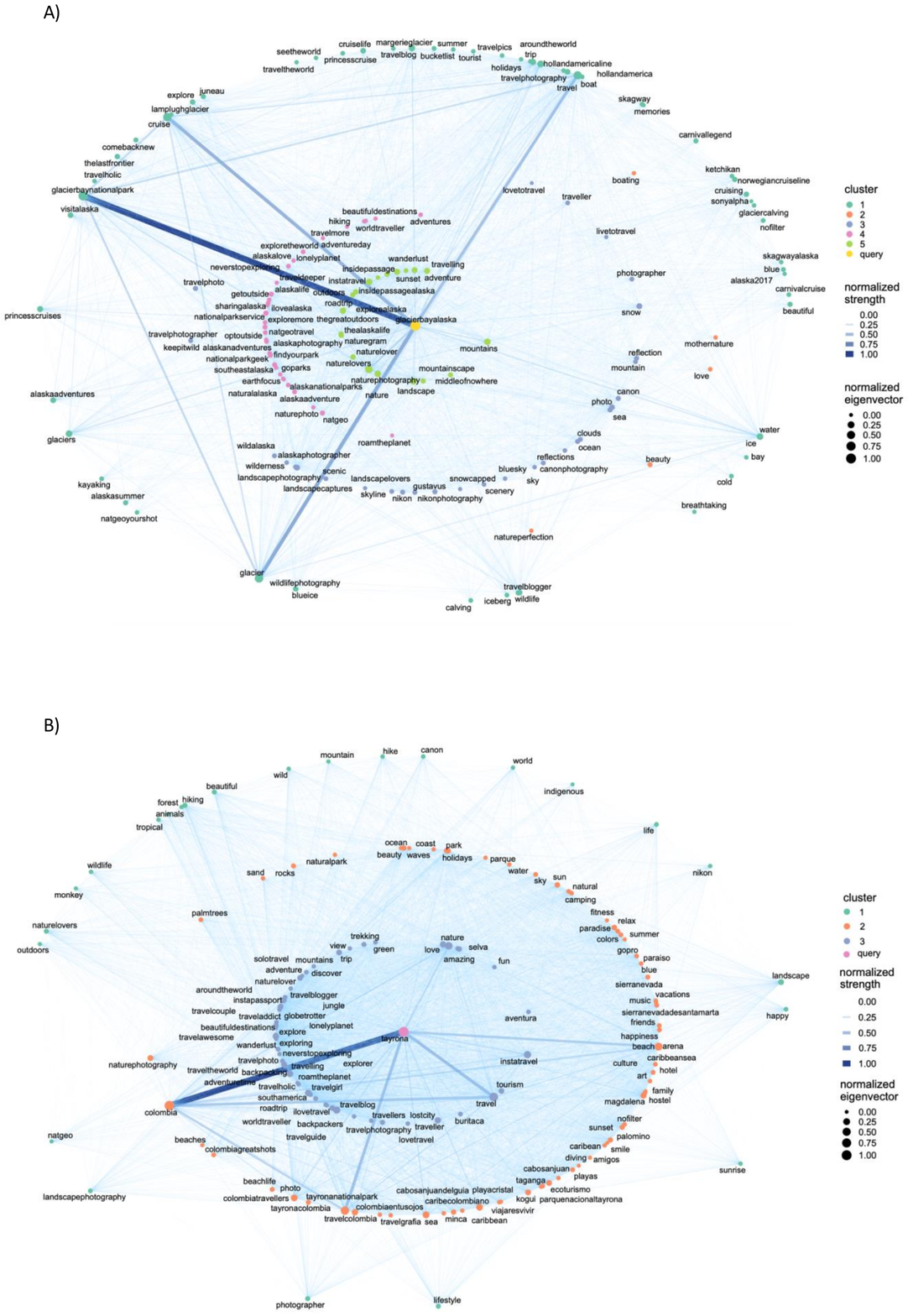}
	\caption{Example of Fast-Greedy community algorithm for the case studies Glacier Bay (A) and Tayrona (B).}
	\label{fig:fig4}
\end{figure}

\begin{longtable}[]{@{}llllll@{}}
\caption{Cultural Ecosystem Services' types depicted from the community analysis (Fast Greedy algorithm). The order of the CES class does not imply a priority rank.}
\label{tab:table2}\\
\hline
\toprule
\endhead
\begin{minipage}[t]{0.14\columnwidth}\raggedright
\strut
\end{minipage} & \begin{minipage}[t]{0.14\columnwidth}\raggedright
CES 1\strut
\end{minipage} & \begin{minipage}[t]{0.14\columnwidth}\raggedright
CES 2\strut
\end{minipage} & \begin{minipage}[t]{0.14\columnwidth}\raggedright
CES 3\strut
\end{minipage} & \begin{minipage}[t]{0.14\columnwidth}\raggedright
CES 4\strut
\end{minipage} & \begin{minipage}[t]{0.14\columnwidth}\raggedright
CES 5\strut
\end{minipage}\tabularnewline
\begin{minipage}[t]{0.14\columnwidth}\raggedright
Galapagos\strut
\end{minipage} & \begin{minipage}[t]{0.14\columnwidth}\raggedright
Nature and wildlife appreciation\strut
\end{minipage} & \begin{minipage}[t]{0.14\columnwidth}\raggedright
Recreational (beach)\strut
\end{minipage} & \begin{minipage}[t]{0.14\columnwidth}\raggedright
Other (travel)\strut
\end{minipage} & \begin{minipage}[t]{0.14\columnwidth}\raggedright
Underwater wildlife and recreational (underwater)\strut
\end{minipage} & \begin{minipage}[t]{0.14\columnwidth}\raggedright
Aesthetic and wellbeing{~}\strut
\end{minipage}\tabularnewline
\begin{minipage}[t]{0.14\columnwidth}\raggedright
Glacier Bay\strut
\end{minipage} & \begin{minipage}[t]{0.14\columnwidth}\raggedright
Aesthetic and nature appreciation\strut
\end{minipage} & \begin{minipage}[t]{0.14\columnwidth}\raggedright
Aesthetic\strut
\end{minipage} & \begin{minipage}[t]{0.14\columnwidth}\raggedright
Recreational (hiking)\strut
\end{minipage} & \begin{minipage}[t]{0.14\columnwidth}\raggedright
Other (National Park and Glaciers)\strut
\end{minipage} & \begin{minipage}[t]{0.14\columnwidth}\raggedright
\strut
\end{minipage}\tabularnewline
\begin{minipage}[t]{0.14\columnwidth}\raggedright
GBR\strut
\end{minipage} & \begin{minipage}[t]{0.14\columnwidth}\raggedright
Underwater wildlife and recreational (underwater)\strut
\end{minipage} & \begin{minipage}[t]{0.14\columnwidth}\raggedright
Other (travel)\strut
\end{minipage} & \begin{minipage}[t]{0.14\columnwidth}\raggedright
Aesthetic and nature appreciation\strut
\end{minipage} & \begin{minipage}[t]{0.14\columnwidth}\raggedright
\strut
\end{minipage} & \begin{minipage}[t]{0.14\columnwidth}\raggedright
\strut
\end{minipage}\tabularnewline
\begin{minipage}[t]{0.14\columnwidth}\raggedright
Isole Egadi\strut
\end{minipage} & \begin{minipage}[t]{0.14\columnwidth}\raggedright
Recreational (water activities)\strut
\end{minipage} & \begin{minipage}[t]{0.14\columnwidth}\raggedright
Aesthetic and wellbeing\strut
\end{minipage} & \begin{minipage}[t]{0.14\columnwidth}\raggedright
Cultural identity\strut
\end{minipage} & \begin{minipage}[t]{0.14\columnwidth}\raggedright
Other (travel)\strut
\end{minipage} & \begin{minipage}[t]{0.14\columnwidth}\raggedright
\strut
\end{minipage}\tabularnewline
\begin{minipage}[t]{0.14\columnwidth}\raggedright
Macquarie Island\strut
\end{minipage} & \begin{minipage}[t]{0.14\columnwidth}\raggedright
Nature and wildlife appreciation\strut
\end{minipage} & \begin{minipage}[t]{0.14\columnwidth}\raggedright
Wildlife and conservation\strut
\end{minipage} & \begin{minipage}[t]{0.14\columnwidth}\raggedright
Recreational and wildlife (iconic fauna)\strut
\end{minipage} & \begin{minipage}[t]{0.14\columnwidth}\raggedright
Wildlife (bird watching)\strut
\end{minipage} & \begin{minipage}[t]{0.14\columnwidth}\raggedright
\strut
\end{minipage}\tabularnewline
\begin{minipage}[t]{0.14\columnwidth}\raggedright
Peninsula Valdez\strut
\end{minipage} & \begin{minipage}[t]{0.14\columnwidth}\raggedright
Wildlife (sea life) and recreation\strut
\end{minipage} & \begin{minipage}[t]{0.14\columnwidth}\raggedright
Wildlife conservation\strut
\end{minipage} & \begin{minipage}[t]{0.14\columnwidth}\raggedright
Aesthetics and recreational\strut
\end{minipage} & \begin{minipage}[t]{0.14\columnwidth}\raggedright
Wildlife (iconic fauna)\strut
\end{minipage} & \begin{minipage}[t]{0.14\columnwidth}\raggedright
\strut
\end{minipage}\tabularnewline
\begin{minipage}[t]{0.14\columnwidth}\raggedright
Easter Island\strut
\end{minipage} & \begin{minipage}[t]{0.14\columnwidth}\raggedright
Cultural heritage\strut
\end{minipage} & \begin{minipage}[t]{0.14\columnwidth}\raggedright
Other (adventure and travel)\strut
\end{minipage} & \begin{minipage}[t]{0.14\columnwidth}\raggedright
Nature, aesthetics and wellbeing\strut
\end{minipage} & \begin{minipage}[t]{0.14\columnwidth}\raggedright
Recreational (underwater)\strut
\end{minipage} & \begin{minipage}[t]{0.14\columnwidth}\raggedright
\strut
\end{minipage}\tabularnewline
\begin{minipage}[t]{0.14\columnwidth}\raggedright
Sandwich Harbour\strut
\end{minipage} & \begin{minipage}[t]{0.14\columnwidth}\raggedright
Aesthetics\strut
\end{minipage} & \begin{minipage}[t]{0.14\columnwidth}\raggedright
Wildlife, aesthetics and recreational\strut
\end{minipage} & \begin{minipage}[t]{0.14\columnwidth}\raggedright
Wellbeing and recreational (safari)\strut
\end{minipage} & \begin{minipage}[t]{0.14\columnwidth}\raggedright
\strut
\end{minipage} & \begin{minipage}[t]{0.14\columnwidth}\raggedright
\strut
\end{minipage}\tabularnewline
\begin{minipage}[t]{0.14\columnwidth}\raggedright
Skomer\strut
\end{minipage} & \begin{minipage}[t]{0.14\columnwidth}\raggedright
Aesthetic and recreation (hiking)\strut
\end{minipage} & \begin{minipage}[t]{0.14\columnwidth}\raggedright
Wildlife (birds) watching\strut
\end{minipage} & \begin{minipage}[t]{0.14\columnwidth}\raggedright
Wildlife (birds)\strut
\end{minipage} & \begin{minipage}[t]{0.14\columnwidth}\raggedright
\strut
\end{minipage} & \begin{minipage}[t]{0.14\columnwidth}\raggedright
\strut
\end{minipage}\tabularnewline
\begin{minipage}[t]{0.14\columnwidth}\raggedright
Tawharanui\strut
\end{minipage} & \begin{minipage}[t]{0.14\columnwidth}\raggedright
Recreational (beach)\strut
\end{minipage} & \begin{minipage}[t]{0.14\columnwidth}\raggedright
Nature, aesthetic and wellbeing\strut
\end{minipage} & \begin{minipage}[t]{0.14\columnwidth}\raggedright
Cultural identity\strut
\end{minipage} & \begin{minipage}[t]{0.14\columnwidth}\raggedright
Wildlife conservation\strut
\end{minipage} & \begin{minipage}[t]{0.14\columnwidth}\raggedright
\strut
\end{minipage}\tabularnewline
\begin{minipage}[t]{0.14\columnwidth}\raggedright
Tayrona\strut
\end{minipage} & \begin{minipage}[t]{0.14\columnwidth}\raggedright
Wellbeing and aesthetics\strut
\end{minipage} & \begin{minipage}[t]{0.14\columnwidth}\raggedright
Recreational (hiking) and cultural heritage\strut
\end{minipage} & \begin{minipage}[t]{0.14\columnwidth}\raggedright
Nature and aesthetics\strut
\end{minipage} & \begin{minipage}[t]{0.14\columnwidth}\raggedright
\strut
\end{minipage} & \begin{minipage}[t]{0.14\columnwidth}\raggedright
\strut
\end{minipage}\tabularnewline
\begin{minipage}[t]{0.14\columnwidth}\raggedright
Togean Island\strut
\end{minipage} & \begin{minipage}[t]{0.14\columnwidth}\raggedright
Other (travel)\strut
\end{minipage} & \begin{minipage}[t]{0.14\columnwidth}\raggedright
Underwater wildlife and recreational (underwater)\strut
\end{minipage} & \begin{minipage}[t]{0.14\columnwidth}\raggedright
Aesthetics, wildlife (underwater) and recreational (underwater)\strut
\end{minipage} & \begin{minipage}[t]{0.14\columnwidth}\raggedright
\strut
\end{minipage} & \begin{minipage}[t]{0.14\columnwidth}\raggedright
\strut
\end{minipage}\tabularnewline
\begin{minipage}[t]{0.14\columnwidth}\raggedright
Vamizi\strut
\end{minipage} & \begin{minipage}[t]{0.14\columnwidth}\raggedright
Nature, wildlife and conservation\strut
\end{minipage} & \begin{minipage}[t]{0.14\columnwidth}\raggedright
Recreational (underwater) and other (luxury tourism)\strut
\end{minipage} & \begin{minipage}[t]{0.14\columnwidth}\raggedright
Aesthetics and wellbeing\strut
\end{minipage} & \begin{minipage}[t]{0.14\columnwidth}\raggedright
Recreational (fishing)\strut
\end{minipage} & \begin{minipage}[t]{0.14\columnwidth}\raggedright
\strut
\end{minipage}\tabularnewline
\begin{minipage}[t]{0.14\columnwidth}\raggedright
Ytrehvaler\strut
\end{minipage} & \begin{minipage}[t]{0.14\columnwidth}\raggedright
Nature and cultural identity\strut
\end{minipage} & \begin{minipage}[t]{0.14\columnwidth}\raggedright
Nature and recreational (hiking and kayak)\strut
\end{minipage} & \begin{minipage}[t]{0.14\columnwidth}\raggedright
Recreational (hiking)\strut
\end{minipage} & \begin{minipage}[t]{0.14\columnwidth}\raggedright
Nature and aesthetics\strut
\end{minipage} & \begin{minipage}[t]{0.14\columnwidth}\raggedright
\strut
\end{minipage}\tabularnewline
\bottomrule
\end{longtable}

\subsection{Merged network of the 14 case studies}
The network that integrates the 14 areas, highlighted several hashtags
that act as bridges between communities of hashtags (Fig. 5).
\emph{Nature}, \emph{travel}, \emph{photo} and \emph{travel photography}
are key to structure the global network. However, the integrated network
evidenced other hashtags with lower eigenvector that also connect
smaller groups. The hashtags \emph{sunset} and \emph{island} connect the
subgroups from Easter Island, Isole Egadi and Vamizi; Tayrona is
connected to this group thought \emph{travelling} and to the central
vertex through \emph{travel photography}.{~}

From this hashtag (\emph{travel photography}) diverges another branch
that connects 7 areas through \emph{adventure}; a small group of
hashtags deriving from this node represent Sandwich harbour and Vamizi,
connected through \emph{Africa}. The hashtag \emph{Ocean}, connected to
\emph{adventure,} relates Great Barrier Reef with Tawharanui, and to
\emph{wanderlust} (a German expression for the desire to explore the
world) that connects Península Valdés, Skomer and Macquairie Island.
These three areas are also connected through the central hashtag
\emph{travel photography}, and Skomer and Macquairie Island
\emph{through wildlife photography}. The hashtag \emph{adventure} is
also connected to a group of hashtags from Galapagos that also derive to
the high eigenvector hashtag \emph{nature}.{~}

The hashtag \emph{nature} is key to include the fragile sub-network
Ytrehvaler, and also derives to other high eigenvector hashtag,
\emph{travel}, that in turn, connects to the small sub-network from
Glacier bay. \emph{Travel} is connected to many less relevant hashtags
that are common to many of the areas: \emph{friends}, \emph{tourist},
\emph{happiness}, and derives to \emph{photo}, another central hashtag.
\emph{Photo} connects to \emph{paradise}, that is key to integrate
Toguean Island, a few hashtags from Tayrona related with the Caribbean
and beach, and a group of hashtags from Peninsula Valdez related with
whale watching. Some other small hashtags, that are connected to high
eigenvector hashtags but are not included in any particular area are
shared by many of the areas, e.g., \emph{sun}, \emph{relax},
\emph{landscape photography}, \emph{nature lovers}, \emph{sunset},
\emph{sky}.{~}

\begin{landscape}
\begin{figure}
	\centering
	\includegraphics[width=1\linewidth]{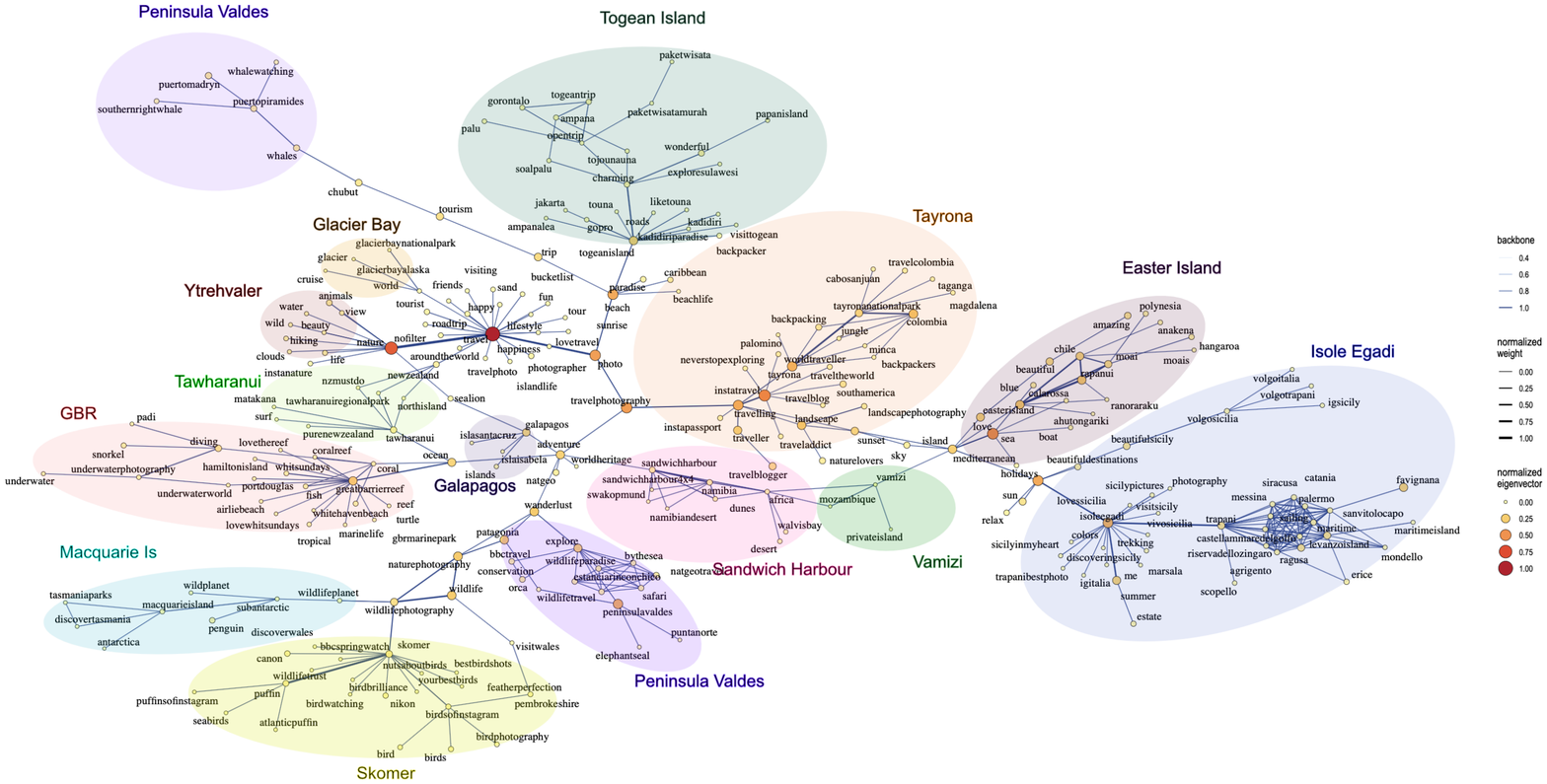}
	\caption{Global network graph including the fourteen case studies where the vertex size represents the Eigenvector centrality and edges represent strength. The coloured clusters arrange the case studies to facilitate the visual identification of related areas}
	\label{fig:fig5}
\end{figure}
\end{landscape}

\section{Discussion}
Social media provides a powerful source of information to monitor
visitors' preferences and perceptions in marine and coastal areas
globally. In our study, the analysis of Instagram data confirmed these
expectations, with underwater activities and underwater life
appreciation prioritized by visitors in iconic diving destinations
(e.g., Great Barrier Reef or Toguean Island), whereas case studies known
to be small natural reserves for wildlife watching were confirmed to be
mainly visited for their iconic fauna (e.g., Península Valdés) or
protected species (e.g., Skomer Island). However, this study differs
from previous studies in the methodology adopted to analyse data stored
in social media platforms. The analysis of photo content is known to be
time consuming and subjected to the researcher's interpretation
{(Ruiz-Frau et al., 2020)}. Conversely, the text linked to each photo
illustrates users' perceptions, preferences, feelings, and emotions. The
analysis of users' posts with Graph Theory allowed the identification of
emergent discourse patterns in Instagram. It identified the most
frequent words related with a specific area (e.g., \emph{nature},
\emph{wildlife} and \emph{photography} in Galapagos), and other less
frequent words connected to the principal ones, delineating a discourse
in each case study, for example information on{ specific knowledge, like
\emph{endemic} and \emph{evolution}, and also feelings, \emph{happy},
related with \emph{wildlife} and \emph{nature} in Galapagos}. The
primary and secondary information provided by the network analysis is of
high value for conservationists and managers, as it delineates visitors
profile and preferences. Importantly, our approach allowed to gather
this individualised information remotely from a {wide variety of marine
and coastal case studies globally.{~}}

It has been acknowledged that the use of information from social media
platforms has an inherent bias associated to both the type of user and
the type of content posted on the platform. {Hausmann et al. (2018)}
observed that while most of the pictures posted on Flickr focused on
biodiversity, Instagram, additionally, was popular for sharing pictures
about activities and people. On the other hand, {Ruiz-Frau et al.
(2020)} observed that Twitter posts reflected social awareness and
discussions around current global concerns such as climate change and
youth movements. Therefore, different social media platforms may be used
by different groups of visitors and, aiming to maximize the
representation of the wider society, we considered Instagram to be a
good candidate for a global assessment of CES covering a wide diversity
of natural spaces. Instagram {provides information on what calls the
attention of visitors, but also on activities and feelings. For example,
many users on Instagram linked natural spaces with conservation (e.g.,
Great Barrier Reef), nature excursions (e.g., Glacier Bay), bird
watching (e.g., Skomer), views of natural fauna (e.g., Península
Valdés), but also, and not least, the scenic beauty (e.g., Sandwich
harbour), the relaxation and happiness provided by open spaces (e.g.,
Galapagos), or some luxury accommodations within national parks as
wellness spaces (e.g., Vamizi).}{ }The main activities reported in each
case study were generally related with the dominant habitat, e.g.,
diving in Great Barrier Reef or Toguean Island, with prevalence of coral
reefs; hiking in Glacier Bay or Tayrona, characterized by forests and
mountains; beach recreation in Tawharanui or Isole Egadi, characterized
by sandy beaches. Nevertheless, the posts' discourse in each area is
markedly defined by the visitor profile, and access to the area. For
example, Vamizi is characterized by luxury and international tourism
that visit the area for the opportunity to enjoy the underwater life,
recreational fishing and beach. Other areas like Skomer and {Ytrehvaler
are mainly visited by locals that enjoy the wildlife and nature of the
place. This implies the CES enjoyed by visitors are highly conditioned
by its logistical accessibility, with remote places like Macquarie
Island or Galapagos visited by international tourists that travel
(\emph{travel}, being one of the most common hashtags) to these areas to
find adventure, recreation, or scenery, amongst others. More accessible
areas (i.e., those easily accessible from large cities), like
Tawharanui, Ytrehvaler or Isole Egadi, appear to be predominantly
visited by locals in search of the relaxation of the beach, sighting of
emblematic local fauna, or nature recreation. The merged network
evidenced that central hashtags to all areas were \emph{nature},
\emph{travel}, \emph{photo}, however, less popular hashtags appear key
to connect smaller groups of areas and were related with general
concepts posted in these areas and not with principal activities or
focus. For example, \emph{sunset} and \emph{island} connected Easter
Island, Isole Egadi and Vamizi, \emph{wanderlust} (a German expression
for the desire to explore the world) connects Península Valdés, Skomer
and Macquairie Island, \emph{paradise} connects Toguean Island and
Tayrona, \emph{ocean} connects Great Barrier Reef and Tawharanui. These
results imply that the perception of visitants to the areas is not
exclusively conditioned by the main activity or ecosystem type, but by
higher level concepts like \emph{paradise} or \emph{wanderlust}.{~}}

What becomes evident from this global assessment is that an area does
not need to be an iconic destination to provide essential services to
society. Galapagos, Great Barrier Reef or Easter Island provide
essential services like nature appreciation, wildlife watching or
cultural identity, however, such services have also been identified in
less iconic areas often visited by local tourists like Skomer Island,
Tawharanui and Ytrehvaler. Similarly, wellbeing related with relax and
happiness is recorded in remote and iconic areas like Galapagos, but
also in quite different places such as Vamizi, Tawharanui or Isole
Egadi. The frequent post of the word \emph{happiness} (and similar
words) denotes the importance of nature's contribution to people's
wellbeing, as emphasised by Russell et al. (2013), Pascual et al.
(2017){ and }Díaz et al. (2018){ that stated that the benefits arising
from human connections with nature include sense of place, identity,
mental health and sense of belonging. These benefits were }independent
of its location, ecosystem or main activity provided. A series of CES
bundles could be identified in each case study, evidencing areas with
high diversity of benefits and perceptions, whereas other areas were
relatively homogenous in users' activities and perceptions. Frequent CES
groups were related with aesthetics and wildlife and nature
appreciation, which is expected as information is obtained from a
photography-based social media platform. However, the classification of
the popular hashtags in CES types, despite providing standardised
information that allows the comparison with other studies, limits the
information provided by the networks. The analysis of social media data
with Graph Theory allowed identification of users' prioritization (e.g.,
landscape, heritage, wildlife), but also activities (e.g., diving,
hiking, relaxing), preferred habitats/species (e.g. forest, beach,
penguins), and feelings (e.g., happiness, beach lifestyle, place
{identity). }Network analysis allowed moving beyond the state of the art
by mere hashtag frequency to the exploration of inter-relations between
hashtags, delineating the users' discourse. For example, animal species
connected to local names provides information on places for wildlife
watching, e.g., puffins in Skomer Island -- \emph{place}, or penguin
species in MacQuarie Island. Hashtags also evidenced environmental
awareness, e.g., conservation in Macquarie Island or climate change in
Great Barrier Reef, which should be considered key to promote
transformative changes for policy makers (Hughes et al., 2018).
Therefore, relevant words like nature watching can be linked to a place
or to a species name, conservation can be linked to a place or ecosystem
component, and so on.

{The methodological approach developed in }{Ruiz-Frau et al. (2020)}{
and used in this work can become an important tool in the assessment of
CES, a key ecosystem services' category that is generally poorly
addressed in management and conservation plans }{(Chan et al., 2016;
IPBES, 2018)}{. However, the use of this methodology is not exempt of
challenges. }Results showed that a manageable sample of posts can
provide valuable information about the CES in a natural area.
Nevertheless, a sufficient volume of posts might not be available for
particular areas. In addition, the application of this methodology is
restricted to those areas with a unique name to be used as a query in
order to avoid downloading information from other areas which might have
the same name. In the present study, we initially explored areas such as
Table Mountain in South Africa or Banc d'Arguin in Mauritania but were
finally discarded as the query downloaded many unrelated posts. {When
using social media data as a proxy for CES there is an inevitable bias
towards aesthetic values }{(Calcagni et al., 2019)}{ and, in the
particular case of Instagram, a strong dominance of content related with
social recreation. Ultimately, photographs tend to express pleasant and
beautiful features }{(Yoshimura and Hiura, 2017)}{ and }{Instagram is
not an exception as in most case studies only positive feelings were
reported, with few exceptions where conservation awareness was
identified in the social media discourse. }{Representativeness can also
be a challenge }{(Guerrero et al., 2016; Tenerelli et al., 2016)}{, and
}{Instagram is mostly representing the younger generations (Abbott et
al., 2013). Perceptions from people that do not post on Instagram,
remarkedly from older generations, }{people without or limited access to
technology or people from countries where Instagram is not sufficiently
dominant are not adequately represented in our approach. Nationality is
also relevant, as CES identified by local visitors can differ from
international tourists }{(Clemente et al., 2019)}{. This can be
partially solved by using words in different languages as queries, like
for Easter Island (e.g., Rapanui, Isla de Pascua and Easter Island), or
by including words in different languages in the network, like in the
Norwegian reserve of }{Ytrehvaler}{. However, social media platforms are
sometimes restricted in certain countries, for example Instagram is not
available in China, or Russian Federation and eastern EU dominant social
network is Vkontakt. This might imply an important bias in the
nationality of tourists encompassed in the analysis. Despite these
limitations, our study approach provides many advantages, including 1)
cost and effort effectiveness, 2) minimization of researchers'
subjectivity, 3) remote collection of information that allows large
scale studies. The assessment of visitors' perceptions in natural spaces
is generally conducted during peak visitation season and restricted to
frequently visited locations }{(Gosal et al., 2019)}{, while the remote
collection of social media data can encompass any temporal dimension,
and, in principle, it covers visitors to all locations within the
natural areas. The variability in visitors' preferences can assist
managers and policy makers design tailored strategies to promote CES
conservation for visitors' enjoyment, which is of high relevance when
destination sites are often ecologically or culturally fragile
}{(Balmford et al., 2004; Ghermandi et al., 2020)}{. The continuous low
cost-effective monitoring of social media can allow a better
understanding of spatial-temporal changes in visitor preferences
(Hausmann et al., 2016), and this approach can now materialise with the
prevalence of smartphones and the posting of experiences in social media
facilitating the remote access to large scale information on peoples'
perceptions and use of natural spaces.}

\section{Conclusions}
It is recognised that effective marine and coastal conservation requires
a large-scale approach and our novel approach has allowed us to collect
data on CES on a wide diversity of marine and coastal areas globally.
\textbf{\emph{}} {The emergent properties of networks of hashtags were
explored to characterise visitors' preferences (e.g., cultural heritage,
wildlife and nature appreciation), but also activities (e.g., diving,
hiking, relaxing), preferred habitats or species (e.g. forest, beach,
penguins), and feelings (e.g., happiness, beach lifestyle, place
}identity). Our approach allowed to identify places valued for their
cultural heritage (e.g., stone statues in Easter Island status), but
also for their iconic species (e.g., puffins in Skomer island) or
natural monuments (e.g., sand dunes in Sandwich harbour), and sense of
place and identity (e.g., Isole Egadi and Tawharanui). Moreover, {the
frequent post of the word \emph{happiness} represents the importance of
nature's contribution to people. }Cultural interactions between humans
and nature are fundamental, including cultural heritage, the iconic
status of certain species or the contributions these make to human
well-being through a sense of place or place identity. The novel
approach introduced here allow to capture these intangible benefits we
obtain from nature in a cost-effective but holistic way, for an
effective management of natural areas, by promoting the integration of
CES into decision making by identifying CES hotspots.{~}

\section*{Acknowledgements}

This work is a product of ECOMAR research network (Evaluation and
monitoring of marine ecosystem services in Iberoamerica; project number
417RT0528, funded by CYTED). Three co-authors were funded by H2020-Marie
Skłodowska-Curie Action during the conduction of this work: SdJ, funded
by MSCA-IF-2016 (ref. 743545); AOA, funded by MSCA-IF-2016 (ref.
746361); ARF, funded by MSCA-IF-2014 (ref. 655475).{~}

\section*{References}

\begin{scriptsize}
Abbott, W., Donaghey, J., Hare, J., Hopkins, P., 2013. An Instagram is
worth a thousand words: an industry panel and audience Q\&A. Library Hi
Tech News 30, 1--6. doi:10.1108/LHTN-08-2013-0047

Balmford, A., Gravestock, P., Hockley, N., McClean, C.J., Roberts, C.M.,
2004. The worldwide costs of marine protected areas. P Natl Acad Sci USA
101, 9694--9697.

Bodin, Ö., Crona, B., Ernstson, H., 2006. Social Networks in Natural
Resource Management: What Is There to Learn from a Structural
Perspective? Ecol Soc 11, resp2. doi:10.5751/ES-01808-1102r02

Bonacich, P., 2007. Some unique properties of eigenvector centrality.
Social Networks 29, 555--564. doi:10.1016/j.socnet.2007.04.002

Brandes, U., Wagner, D., 2004. Analysis and Visualization of Social
Networks, in: Untangling the Hairballs of Multi-Centered, Small-World
Online Social Media Networks. Springer, Berlin, Heidelberg, Berlin,
Heidelberg, pp. 321--340. doi:10.1007/978-3-642-18638-7\_15

Calcagni, F., Maia, A.T.A., Connolly, J.J.T., Langemeyer, J., 2019.
Digital co-construction of relational values: understanding the role of
social media for sustainability. Sustain Sci 14, 1309--1321.
doi:10.1007/s11625-019-00672-1

Chan, K.M.A., Balvanera, P., Benessaiah, K., Chapman, M., Díaz, S.,
Gómez-Baggethun, E., Gould, R., Hannahs, N., Jax, K., Klain, S., Luck,
G.W., Martín-López, B., Muraca, B., Norton, B., Ott, K., Pascual, U.,
Satterfield, T., Tadaki, M., Taggart, J., Turner, N., 2016. Opinion: Why
protect nature? Rethinking values and the environment 113, 1462--1465.
doi:10.1073/pnas.1525002113

Chan, K.M.A., Satterfield, T., Goldstein, J., 2012. Rethinking ecosystem
services to better address and navigate cultural values. Ecol Econ 74,
8--18. doi:10.1016/j.ecolecon.2011.11.011

Chen, W., Van Assche, K.A.M., Hynes, S., Bekkby, T., Christie, H.C.,
Gundersen, H., 2020. Ecosystem accounting's potential to support coastal
and marine governance. Mar Policy 112, 103758.
doi:10.1016/j.marpol.2019.103758

Clemente, P., Calvache, M., Antunes, P., Santos, R., Cerdeira, J.O.,
Martins, M.J., 2019. Combining social media photographs and species
distribution models to map cultural ecosystem services: The case of a
Natural Park in Portugal. Ecol. Ind. 96, 59--68.

Costanza, R., de Groot, R., Sutton, P., van der Ploeg, S., Anderson,
S.J., Kubiszewski, I., Farber, S., Turner, R.K., 2014. Changes in the
global value of ecosystem services. Global Environ Chang 26, 152--158.
doi:10.1016/j.gloenvcha.2014.04.002

Csardi, G., Nepusz, T., 2006. The igraph software package for complex
network research. Interjournal Complex Sy, 1695.

Daniel, T.C., Muhar, A., Arnberger, A., Aznar, O., Boyd, J.W., Chan,
K.M.A., Costanza, R., Elmqvist, T., Flint, C.G., Gobster, P.H.,
Grêt-Regamey, A., Lave, R., Muhar, S., Penker, M., Ribe, R.G.,
Schauppenlehner, T., Sikor, T., Soloviy, I., Spierenburg, M.,
Taczanowska, K., Tam, J., Dunk, von der, A., 2012. Contributions of
cultural services to the ecosystem services agenda. P Natl Acad Sci USA
109, 8812--8819. doi:10.1073/pnas.1114773109

Di Minin, E., Fink, C., Tenkanen, H., Hiippala, T., 2018. Machine
learning for tracking illegal wildlife trade on social media. Nat Ecol
Evol 2, 406--407. doi:10.1038/s41559-018-0466-x

Díaz, S., Pascual, U., Stenseke, M., Martín-López, B., Watson, R.T.,
Molnár, Z., Hill, R., Chan, K.M.A., Baste, I.A., Brauman, K.A., Polasky,
S., Church, A., Lonsdale, M., Larigauderie, A., Leadley, P.W., van
Oudenhoven, A.P.E., van der Plaat, F., Schröter, M., Lavorel, S.,
Aumeeruddy-Thomas, Y., Bukvareva, E., Davies, K., Demissew, S., Erpul,
G., Failler, P., Guerra, C.A., Hewitt, C.L., Keune, H., Lindley, S.,
Shirayama, Y., 2018. Assessing nature's contributions to people. Science
359, 270--272. doi:10.1126/science.aap8826

Figueroa-Alfaro, R.W., Tang, Z., 2016. Evaluating the aesthetic value of
cultural ecosystem services by mapping geo-tagged photographs from
social media data on Panoramio and Flickr. Journal of Environmental
Planning and Management 60, 266--281. doi:10.1080/09640568.2016.1151772

Freeman, L.C., 1979. Centrality in Social Networks Conceptual
Clarification. Social Networks 215--239.

Geboers, M.A., Van De Wiele, C.T., 2020. Machine Vision and Social Media
Images: Why Hashtags Matter. Social Media + Society 6, 205630512092848.
doi:10.1177/2056305120928485

Ghermandi, A., Camacho-Valdez, V., Trejo-Espinosa, H., 2020. Social
media-based analysis of cultural ecosystem services and heritage tourism
in a coastal region of Mexico. Tourism Management 77, 104002.

Girvan, M., Newman, M.E.J., 2002. Community structure in social and
biological networks. P Natl Acad Sci USA 99, 7821--7826.
doi:10.1073/pnas.122653799

Gosal, A.S., Geijzendorffer, I.R., Václavík, T., Poulin, B., Ziv, G.,
2019. Using social media, machine learning and natural language
processing to map multiple recreational beneficiaries. Ecosystem
Services 38, 100958. doi:10.1016/j.ecoser.2019.100958

Guerrero, P., Møller, M.S., Olafsson, A.S., Snizek, B., 2016. Revealing
Cultural Ecosystem Services through Instagram Images: The Potential of
Social Media Volunteered Geographic Information for Urban Green
Infrastructure Planning and Governance. Urban Planning 1, 1--17.
doi:10.17645/up.v1i2.609

Haines-Young, R., Potschin, M., 2010. The links between biodiversity,
ecosystem services and human well-being, in: Raffaelli, D., Frid, C.L.
(Eds.), Ecosystem Ecology: a New Synthesis. BES Ecological Reviews
Series, Cambridge, pp. 110--139.

Hale, R.L., Cook, E.M., Beltrán, B.J., 2019. Cultural ecosystem services
provided by rivers across diverse social-ecological landscapes: A social
media analysis. Ecol. Ind. 107, 105580.
doi:10.1016/j.ecolind.2019.105580

Hausmann, A., Toivonen, T., Slotow, R., Tenkanen, H., Moilanen, A.,
Heikinheimo, V., Di Minin, E., 2018. Social Media Data Can Be Used to
Understand Tourists' Preferences for Nature‐Based Experiences in
Protected Areas. Consrv Let 11, e12343. doi:10.1111/conl.12343

Hughes, T.P., Kerry, J.T., Baird, A.H., Connolly, S.R., Dietzel, A.,
Eakin, C.M., Heron, S.F., Hoey, A.S., Hoogenboom, M.O., Liu, G.,
McWilliam, M.J., Pears, R.J., Pratchett, M.S., Skirving, W.J., Stella,
J.S., Torda, G., 2018. Global warming transforms coral reef assemblages.
Nature 556, 492--496. doi:10.1038/s41586-018-0041-2

IPBES, 2018. Summary for policymakers of the regional assessment report
on biodiversity and ecosystem services for Europe and Central Asia of
the Intergovernmental Science-Policy Platform on Biodiversity and
Ecosystem Services. IPBES secretariat, Bonn, Germany.

Jeawak, S.S., Jones, C.B., Schockaert, S., 2017. Using Flickr for
Characterizing the Environment: An Exploratory Analysis. DROPS-IDN/7752
13. doi:10.4230/LIPIcs.COSIT.2017.21

Kirchhoff, T., 2012. Pivotal cultural values of nature cannot be
integrated into the ecosystem services framework. P Natl Acad Sci USA
109, E3146--E3146. doi:10.1073/pnas.1212409109

Klain, S.C., Satterfield, T.A., Chan, K.M.A., 2014. What matters and
why? Ecosystem services and their bundled qualities. Ecol Econ 107,
310--320. doi:10.1016/j.ecolecon.2014.09.003

Lee, H., Seo, B., Koellner, T., Lautenbach, S., 2019. Mapping cultural
ecosystem services 2.0 -- Potential and shortcomings from unlabeled
crowd sourced images. Ecol. Ind. 96, 505--515.
doi:10.1016/j.ecolind.2018.08.035

Liquete, C., Piroddi, C., Drakou, E.G., Gurney, L., Katsanevakis, S.,
Charef, A., Egoh, B., 2013. Current Status and Future Prospects for the
Assessment of Marine and Coastal Ecosystem Services: A Systematic
Review. PloS ONE 8, e67737. doi:10.1371/journal.pone.0067737

Maiya, A.S., Berger-Wolf, T.Y., 2010. Online Sampling of High Centrality
Individuals in Social Networks, in: Advances in Knowledge Discovery and
Data Mining, 14th Pacific-Asia Conference, PAKDD 2010, Hyderabad, India,
June 21-24, 2010. Proceedings. Part I. Springer, Berlin, Heidelberg,
Berlin, Heidelberg, pp. 91--98. doi:10.1007/978-3-642-13657-3\_12

Milcu, A.I., Hanspach, J., Abson, D., Fischer, J., 2013. Cultural
Ecosystem Services: a literature review and prospects for future
research. Ecology and Society 18, 44. doi:10.2307/26269377

Newman, M.E.J., 2004. Fast algorithm for detecting community structure
in networks. Phys. Rev. E 69, 066133. doi:10.1103/PhysRevE.69.066133

Nocaj, A., Ortmann, M., Brandes, U., 2015. Untangling the Hairballs of
Multi-Centered, Small-World Online Social Media Networks. JGAA 19,
595--618. doi:10.7155/jgaa.00370

Oteros-Rozas, E., Martín-López, B., Fagerholm, N., Bieling, C.,
Plieninger, T., 2018. Using social media photos to explore the relation
between cultural ecosystem services and landscape features across five
European sites. Ecol. Ind. 94, 74--86. doi:10.1016/j.ecolind.2017.02.009

Oteros-Rozas, E., Martín-López, B., González, J.A., Plieninger, T.,
López, C.A., Montes, C., 2014. Socio-cultural valuation of ecosystem
services in a transhumance social-ecological network. Reg Environ Change
14, 1269--1289. doi:10.1007/s10113-013-0571-y

Pascual, U., Balvanera, P., Díaz, S., Pataki, G., Roth, E., Stenseke,
M., Watson, R.T., Başak Dessane, E., Islar, M., Kelemen, E., Maris, V.,
Quaas, M., Subramanian, S.M., Wittmer, H., Adlan, A., Ahn, S.,
Al-Hafedh, Y.S., Amankwah, E., Asah, S.T., Berry, P., Bilgin, A.,
Breslow, S.J., Bullock, C., Cáceres, D., Daly-Hassen, H., Figueroa, E.,
Golden, C.D., Gómez-Baggethun, E., González-Jiménez, D., Houdet, J.,
Keune, H., Kumar, R., Ma, K., May, P.H., Mead, A., O'Farrell, P.,
Pandit, R., Pengue, W., Pichis-Madruga, R., Popa, F., Preston, S.,
Pacheco-Balanza, D., Saarikoski, H., Strassburg, B.B., van den Belt, M.,
Verma, M., Wickson, F., Yagi, N., 2017. Valuing nature's contributions
to people: the IPBES approach. Current Opinion in Environmental
Sustainability 26-27, 7--16. doi:10.1016/j.cosust.2016.12.006

Plieninger, T., Dijks, S., Oteros-Rozas, E., Bieling, C., 2013.
Assessing, mapping, and quantifying cultural ecosystem services at
community level. Land Use Policy 33, 118--129.
doi:10.1016/j.landusepol.2012.12.013

R Core Team (2019) R: A Language and Environment for Statistical
Computing. Vienna, Austria

Retka, J., Jepson, P., Ladle, R.J., Malhado, A.C.M., Vieira, F.A.S.,
Normande, I.C., Souza, C.N., Bragagnolo, C., Correia, R.A., 2019.
Assessing cultural ecosystem services of a large marine protected area
through social media photographs. Ocean Coast Manage 176, 40--48.
doi:10.1016/j.ocecoaman.2019.04.018

Rodrigues, J.G., Conides, A., Rodriguez, S.R., Raicevich, S., Pita, P.,
Kleisner, K., Pita, C., Lopes, P., Roldán, V.A., Ramos, S., Klaoudatos,
D., Outeiro, L., Armstrong, C., Teneva, L., Stefanski, S.,
Böhnke-Henrichs, A., Kruse, M., Lillebø, A., Bennett, E., Belgrano, A.,
Murillas, A., Pinto, I.S., Burkhard, B., Villasante, S., 2017. Marine
and Coastal Cultural Ecosystem Services: knowledge gaps and research
priorities. Marine and Coastal Cultural Ecosystem Services: knowledge
gaps and research priorities 2, e12290. doi:10.3897/oneeco.2.e12290

Roth, C., Cointet, J.-P., 2010. Social and semantic coevolution in
knowledge networks. Social Networks 32, 16--29.
doi:10.1016/j.socnet.2009.04.005

Ruiz-Frau, A., Edwards-Jones, G., Kaiser, M.J., 2011. Mapping
stakeholder values for coastal zone management. Mar Ecol Prog Ser 434,
239--249. doi:10.3354/meps09136

Ruiz-Frau, A., Ospina-Alvarez, A., Villasante, S., Pita, P.,
Maya-Jariego, I., Mohan, S. de J., 2020. Using graph theory and social
media data to assess cultural ecosystem services in coastal areas:
Method development and application.

Russell, R., Guerry, A.D., Balvanera, P., Gould, R.K., Basurto, X.,
Chan, K.M.A., Klain, S., Levine, J., Tam, J., 2013. Humans and Nature:
How Knowing and Experiencing Nature Affect Well-Being.
http://dx.doi.org/10.1146/annurev-environ-012312-110838 38, 473--502.
doi:10.1146/annurev-environ-012312-110838

Spalding, M.D., Fox, H.E., Allen, G.R., Davidson, N., Ferdaña, Z.A.,
Finlayson, M., Halpern, B.S., Jorge, M.A., Lombana, A., Lourie, S.A.,
Martin, K.D., McManus, E., Molnar, J., Recchia, C.A., Robertson, J.,
2007. Marine Ecoregions of the World: A Bioregionalization of Coastal
and Shelf Areas. BioScience 57, 573--583. doi:10.1641/B570707

Tenerelli, P., Demšar, U., Luque, S., 2016. Crowdsourcing indicators for
cultural ecosystem services: A geographically weighted approach for
mountain landscapes. Ecol. Ind. 64, 237--248.
doi:10.1016/j.ecolind.2015.12.042

Teoh, S.H.S., Symes, W.S., Sun, H., Pienkowski, T., Carrasco, L.R.,
2019. A global meta-analysis of the economic values of provisioning and
cultural ecosystem services. Sci Total Environ 649, 1293--1298.
doi:10.1016/j.scitotenv.2018.08.422

Topirceanu, A., Udrescu, M., Marculescu, R., 2018. Weighted Betweenness
Preferential Attachment: A New Mechanism Explaining Social Network
Formation and Evolution. Sci. Rep. 8, 1--14.
doi:10.1038/s41598-018-29224-w

Varol, O., Ferrara, E., Davis, C.A., Menczer, F., Flammini, A., 2017.
Online Human-Bot Interactions: Detection, Estimation, and
Characterization. Eleventh International AAAI Conference on Web and
Social Media.

Vaz, A.S., Gonçalves, J.F., Pereira, P., Santarém, F., Vicente, J.R.,
Honrado, J.P., 2019. Earth observation and social media: Evaluating the
spatiotemporal contribution of non-native trees to cultural ecosystem
services. Remote Sensing of Environment 230, 111193.
doi:10.1016/j.rse.2019.05.012

Yoshimura, N., Hiura, T., 2017. Demand and supply of cultural ecosystem
services: Use of geotagged photos to map the aesthetic value of
landscapes in Hokkaido. Ecosystem Services 24, 68--78.
doi:10.1016/j.ecoser.2017.02.00

\end{scriptsize}

\appendix
\label{sec:Appendix}
\includepdf[pages=1-13]{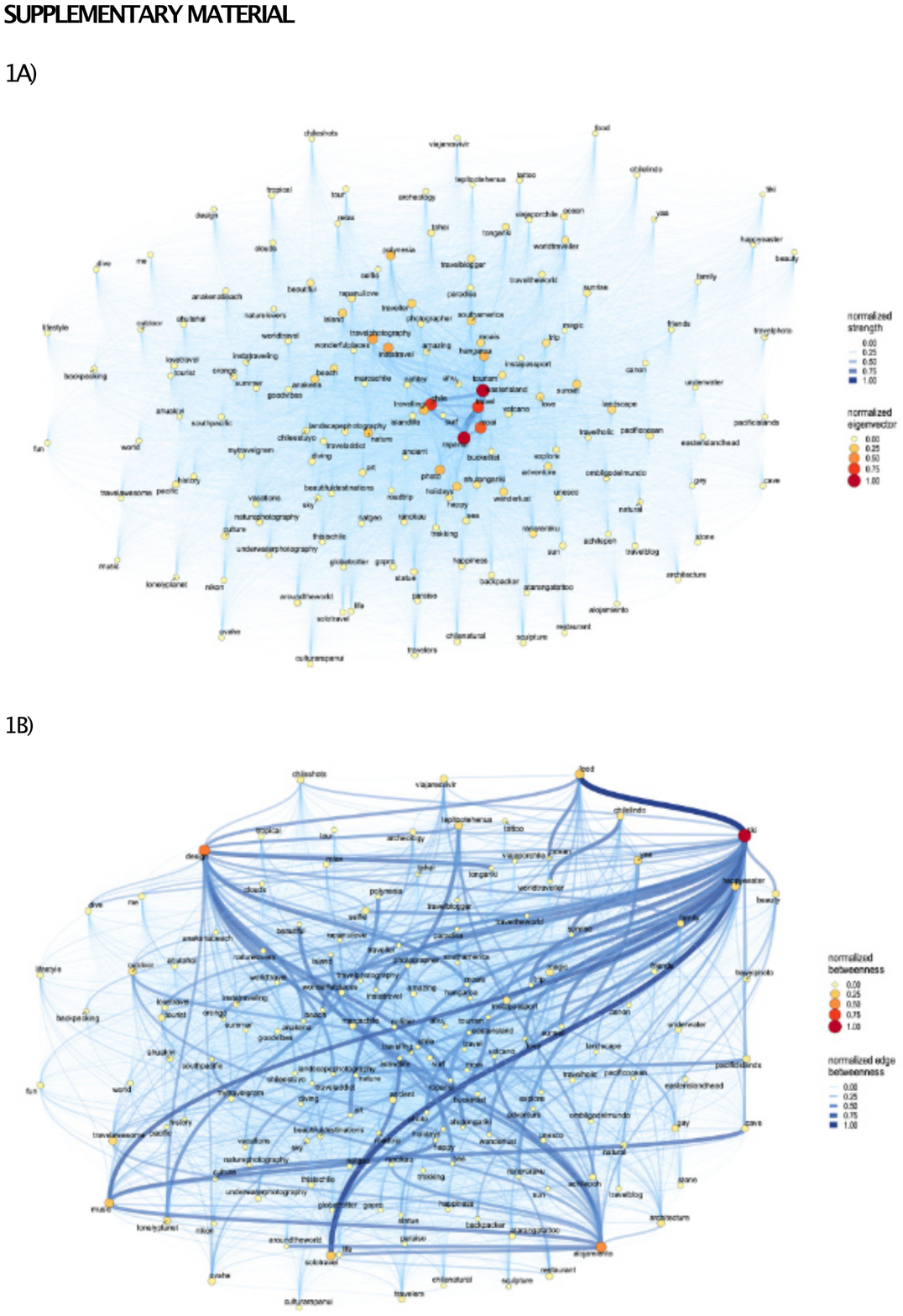}
\includepdf[pages=1-6]{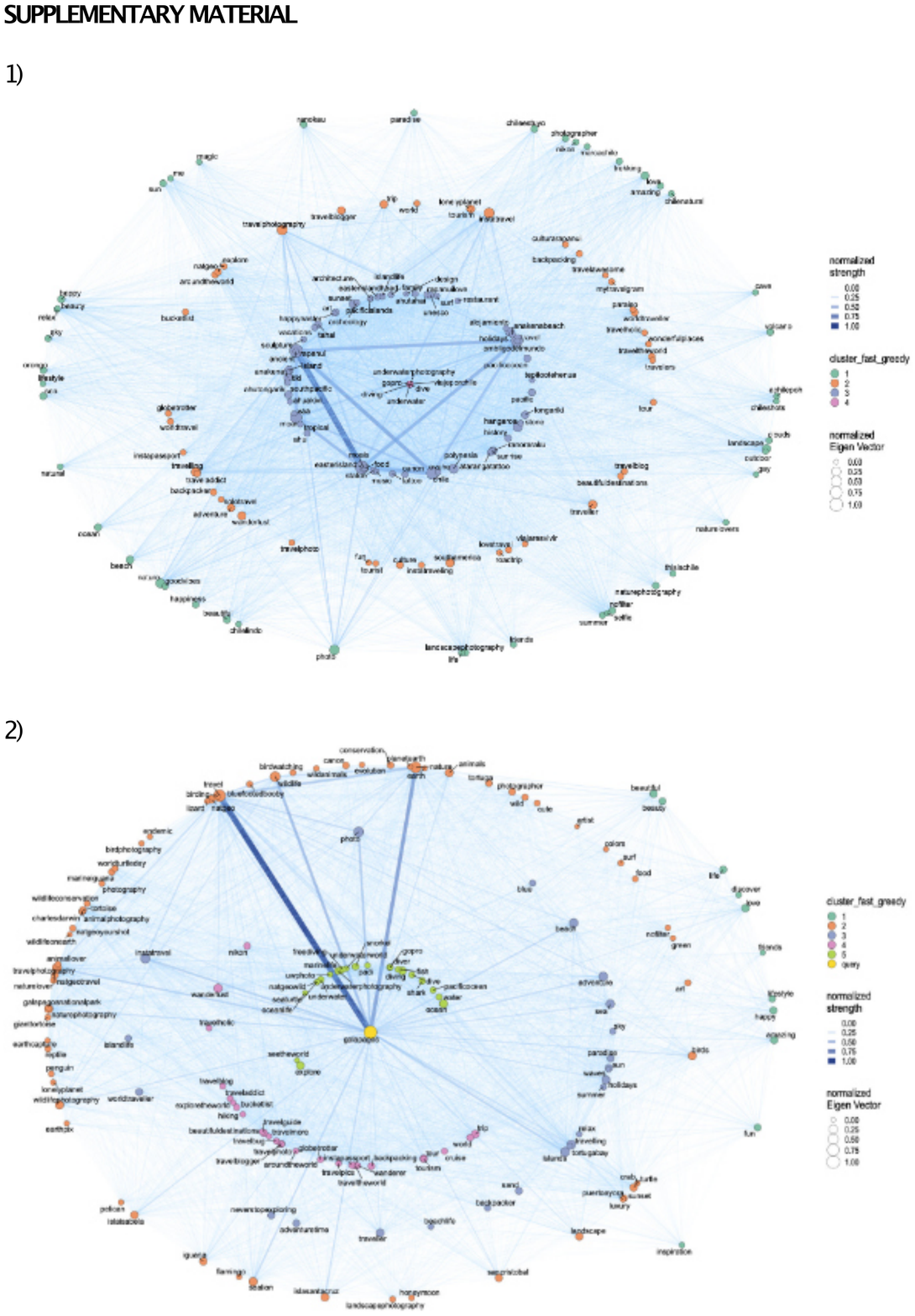}
\includepdf[pages=1]{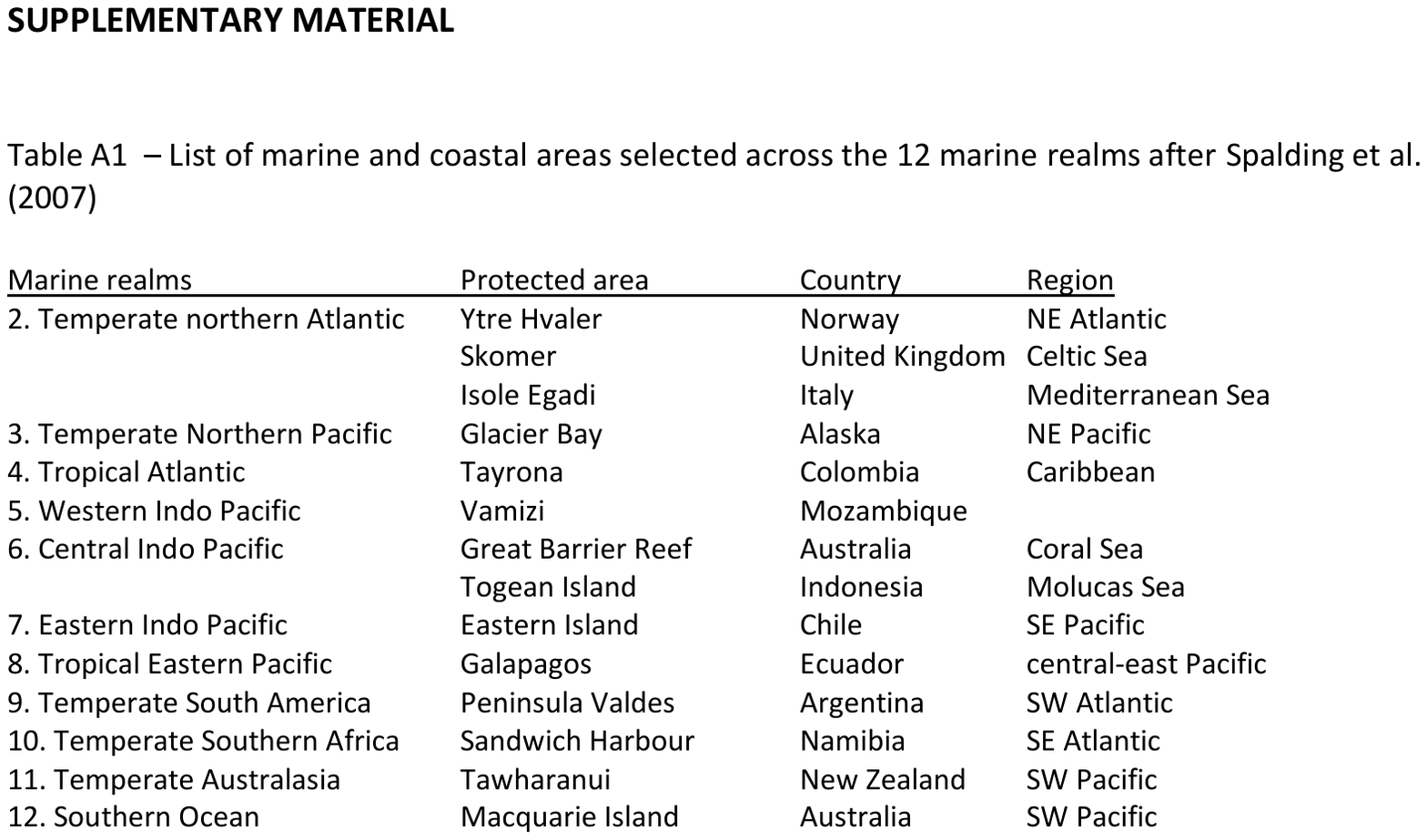}

\end{document}